\documentclass[preprint, review, 12pt, 3p]{elsarticle}
\usepackage{amsmath}
\usepackage{scrextend}
\usepackage{rotating}
\usepackage{graphicx}
\usepackage{enumitem}
\usepackage{subfig}
\usepackage{dirtytalk}
\usepackage{multirow}
\usepackage[utf8]{inputenc}
\usepackage{lipsum}
\usepackage{caption}
\usepackage{adjustbox}
\usepackage{tabularx}
\usepackage{floatrow} 
\usepackage[vlines]{tabularht}
\usepackage{pbox}
\usepackage{float}
\floatstyle{plaintop}
\restylefloat{table}

\newcommand{\concat}{\ensuremath{+\!\!\!\!+\,}}

\usepackage{url}

\usepackage{soul}
\usepackage[linesnumbered,ruled,vlined]{algorithm2e}
\SetKwInput{KwInput}{Input}                
\SetKwInput{KwOutput}{Output}              

\usepackage[table,xcdraw]{xcolor}

\newsavebox\MBox

\journal{Biomedical Signal Processing and Control}

\begin{document}

\begin{frontmatter}

\title{Dermo-DOCTOR: A framework for concurrent skin lesion detection and recognition using a deep convolutional neural network with end-to-end dual encoders}

\author[label1]{Md. Kamrul Hasan\corref{cor1}\fnref{label3}}
\address[label1]{Department of Electrical and Electronic Engineering, Khulna University of Engineering \& Technology, Khulna-9203, Bangladesh}
\address[label7]{Department of Biomedical Engineering, Khulna University of Engineering \& Technology, Khulna-9203, Bangladesh}
\address[label6]{Department of Computer Science and Engineering, Khulna University of Engineering \& Technology, Khulna-9203, Bangladesh}
\cortext[cor1]{I am corresponding author}
\fntext[label4]{Department of EEE, KUET, Khulna-9203, Bangladesh.}
\ead{m.k.hasan@eee.kuet.ac.bd}
\author[label1]{Shidhartho Roy}
\ead{swapno15roy@gmail.com}
\author[label1]{Chayan Mondal}
\ead{chayan.eee.92@gmail.com}
\author[label1]{Md. Ashraful Alam}
\ead{ashrafulalam16e@gmail.com}
\author[label1]{Md. Toufick E Elahi}
\ead{toufick1469@gmail.com}
\author[label7]{Aishwariya Dutta}
\ead{aishwariyadutta16@gmail.com}
\author[label6]{S. M. Taslim Uddin Raju}
\ead{taslimuddinraju7864@gmail.com}
\author[label1]{Md. Tasnim Jawad}
\ead{jawad1703006@stud.kuet.ac.bd}
\author[label1]{Mohiuddin Ahmad}
\ead{ahmad@eee.kuet.ac.bd}

\begin{abstract}

\subsection*{Background and Objective}
Automated skin lesion analysis for simultaneous detection and recognition is still challenging for inter-class homogeneity and intra-class heterogeneity, leading to low generic capability of a Single Convolutional Neural Network (CNN) with limited datasets.

\subsection*{Methods} 
This article proposes an end-to-end deep CNN-based framework for simultaneous detection and recognition of the skin lesions, named Dermo-DOCTOR, consisting of two encoders. The feature maps from two encoders are fused channel-wise, called Fused Feature Map (FFM).
The FFM is utilized for decoding in the detection sub-network, concatenating each stage of two encoders' outputs with corresponding decoder layers to retrieve the lost spatial information due to pooling in the encoders. 
For the recognition sub-network, the outputs of three fully connected layers, utilizing feature maps of two encoders and FFM, are aggregated to obtain a final lesion class. 
We train and evaluate the proposed Dermo-Doctor utilizing two publicly available benchmark datasets, such as ISIC-$2016$ and ISIC-$2017$. 

\subsection*{Results} 
The achieved segmentation results exhibit mean intersection over unions of $85.0\,\%$ and $80.0\,\%$ respectively for ISIC-$2016$ and ISIC-$2017$ test datasets. The proposed Dermo-DOCTOR also demonstrates praiseworthy success in lesion recognition, providing the areas under the receiver operating characteristic curves of $0.98$ and $0.91$ respectively for those two datasets. The experimental results show that the proposed Dermo-DOCTOR outperforms the alternative methods mentioned in the literature, designed for skin lesion detection and recognition.

\subsection*{Conclusion}
As the Dermo-DOCTOR provides better-results on two different test datasets, even with limited training data, it can be an auspicious computer-aided assistive tool for dermatologists. 

\end{abstract}

\begin{keyword}
Malignant melanoma \sep Skin lesion detection and recognition \sep Convolutional neural networks \sep Dual encoder networks \sep ISIC skin lesion datasets. 
\end{keyword}

\end{frontmatter}

\section{Introduction}
\label{introduction}
\subsection{Problem Presentation}
\begin{figure*}[!ht]
  \centering
  \subfloat{\includegraphics[width=15cm, height= 9cm]{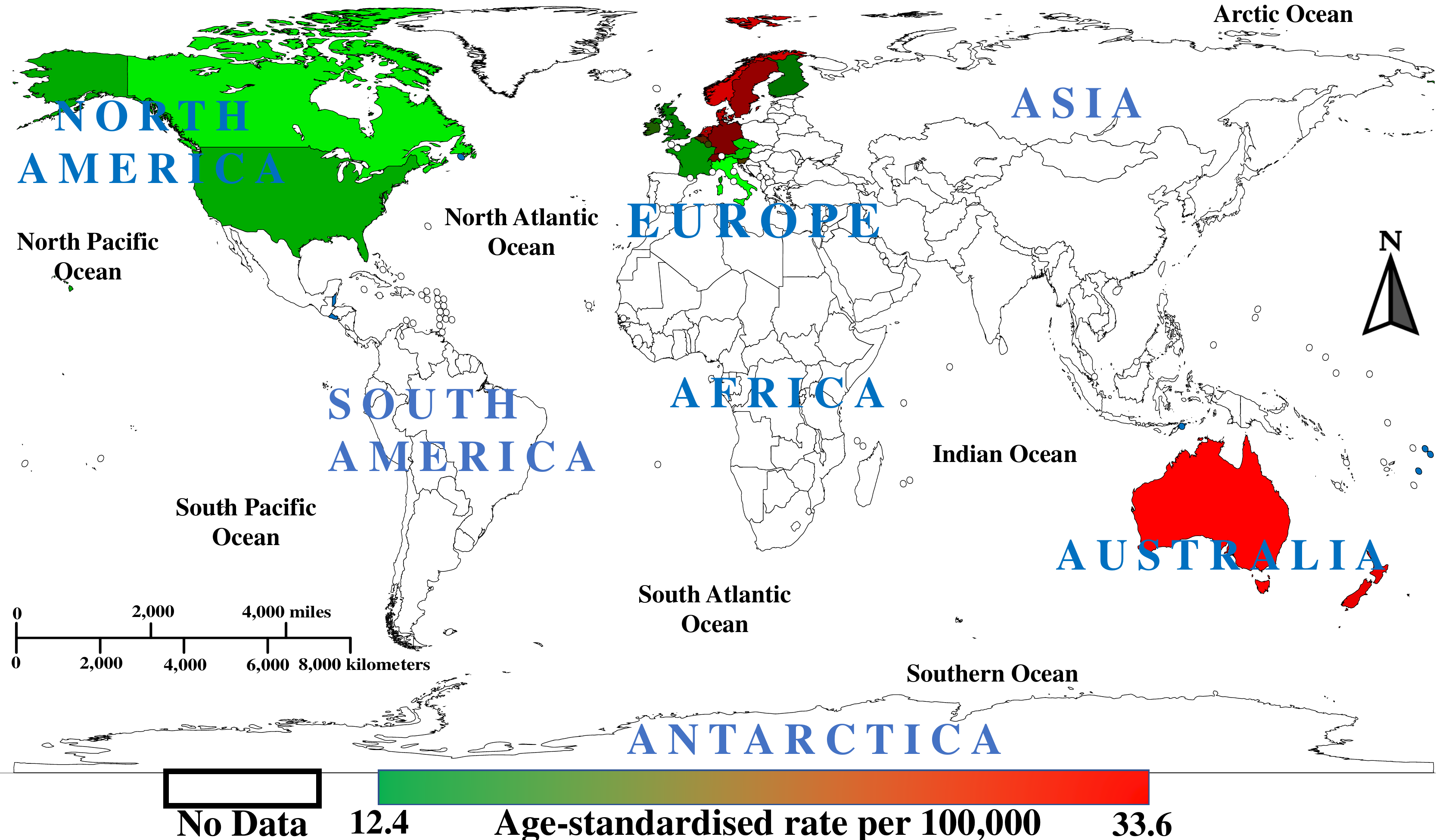}} 
  \caption{A world heat map of the age-standardized rates per $1.0$ million population of the top $20$ countries with the highest skin melanoma rates in $2018$ \citep{bray2018global}, displaying highly skin cancer affected regions.}
  \label{fig:Staistics}
\end{figure*}

\begin{figure*}[!ht]
  \centering
  \subfloat{\includegraphics[width=16cm, height=4.9cm]{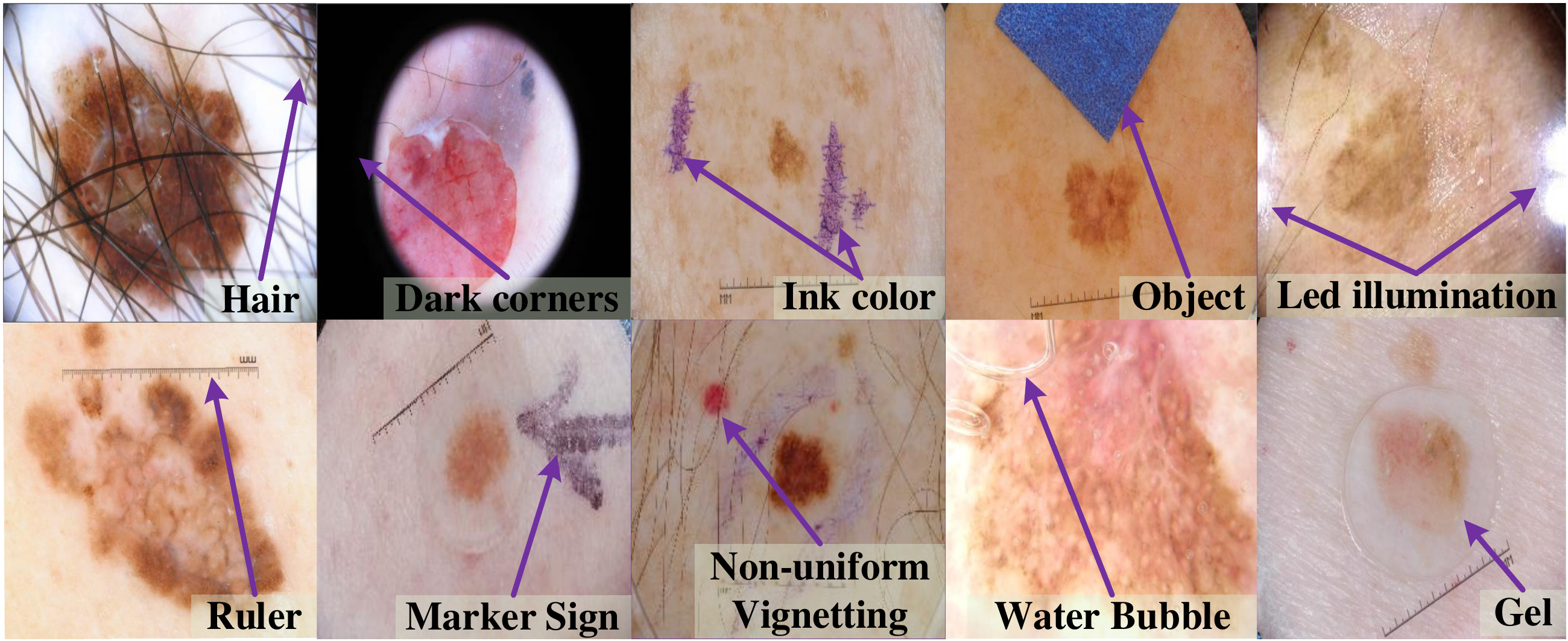}} 
  \caption{An example of the challenging dermoscopic images in ISIC dataset \citep{codella2018skin,gutman2016skin} with different artifacts \citep{hasan2020dsnet}.}
  \label{fig:artifacts}
\end{figure*}

Cancer is an abnormal and uncontrolled growth of dividing cells, damaging different body cells and contributing to the world's second-leading cause of death \citep{siegel2020cancer}.
Although melanomas constitute less than $5.0\,\%$ of all skin cancers, they make up about $75.0\,\%$ of deaths related to skin cancer in the United States (US) alone \citep{estava2017dermatologist}.
Age-standardized melanoma rates of the top $20$ countries \citep{bray2018global} is presented in Fig.~\ref{fig:Staistics}.
According to the world health organization, $466914$ new cases of skin cancer will be diagnosed in $2040$ ($54.27\,\%$ men and $45.73\,\%$ women), and $105904$ ($58.14\,\%$ men and $41.86\,\%$ women) will die.
The five-year survival rate of melanoma, the deadliest variety of early detection, is as high as $99.0\,\%$, but delayed diagnosis leads significantly to a colorectal survival rate decrease of $23.0\,\%$ \citep{venugopal2019colorectal}. 
However, reliable early identification is highly imperative as the five-year survival rate will be increased by $90.0\,\%$ approximately \citep{ge2017skin}. 
Dermatologists generally examine images via naked-eye through visual examination, requiring a high level of expertise and focus. The manual inspection by dermatologists is often very tiresome, time-consuming, subjective, and fault-prone. The precision of skin lesions' diagnosis by the dermatologists suffers from inter-class homogeneity and intra-class heterogeneity. Moreover, the ratio of dermatologists per $1.0$ million population in the US, South Australia, and Europe are respectively $34.0$, $26.0$, and $59.34$, which are very low compared to the required numbers \citep{Europen17online,Dermatology16Factsheet,glazer2017analysis}. 
However, an automated Computer-aided Screening (CAS) system has become popular among dermatologists to alleviate the above limitations, reduce the working burden of dermatologists, and accelerate diagnosis rates \citep{lattoofi2019melanoma}. 
Such CAS systems essentially consist of several integral parts, where the segmentation for Region of Interest (ROI) extraction and the classification for lesion recognition. However, image-based automated CAS systems are highly challenging for the following hurdles: 
\begin{itemize}
    \item Wide range of intra-class variance in colors, textures, edges, and shapes and homogeneity in inter-classes.
     
    \item Sometimes, low contrasts and unclear boundaries (edges) in the malignant and other class images.
     
    \item Lesion ROI frequently shares similar visual characteristics and subtle distinctions due to lighting, perspective, and spatial information within an image.
     
    \item The appearance of different artifacts, such as natural (hairs, veins) or synthetic (air bubbles, ruler lines, color balance charts, marker signs, paint, ink color, artificial objects, \textit{etc.}), LED lighting, darker border (microscopic effects), and non-uniform vignetting, as depicted in Fig.~\ref{fig:artifacts}.

    \item Lesion ROI only covers a small proportion of local, subtle grain, and global context information.
    
    \item Unavailability of a large number of manually annotated images, which is the core requirement of the supervised learning systems.
\end{itemize}

\subsection{Recent Methods}
\label{Methods}
The state-of-the-art methods for skin lesion segmentation and recognition are reviewed and described in the following two subsections. 
\subsubsection{Methods for Lesion Segmentation}
A multi-stage Fully Convolutional Network (mFCN) with parallel integration was introduced by \citet{bi2017dermoscopic}. During the training process, mFCN learned from both the training data and the coarse results obtained from the previous (m-1)FCN stage. The summation of the earlier results with the current result had two benefits: boost the training data and optimize lesion boundary learning. 
\citet{navarro2018accurate} proposed a superpixels adaptation-based segmentation approach to get tight-to-boundaries of the skin lesions. 
The authors applied the Scale-Invariant Feature Transform (SIFT) \citep{lowe2004distinctive} and Gaussian distribution to detect the feature points and place these points to the initial centers. 
Finally, they applied the simple linear iterative clustering technique to these points for generating the final lesion masks.
An encoder-decoder network was built by \citet{sarker2018slsdeep}, called SLSDeep. The encoder in SLSDeep was dilated residual network-based design, while the decoder had a pyramid pooling network. Additionally, they proposed a combined negative log-likelihood and endpoint error-based cross-entropy loss function. 
\citet{jahanifar2018supervised} developed an improved saliency detection supervised method for the lesion segmentation, which was designed based on the discriminative regional feature integration. They used a thresholding algorithm for generating a new pseudo background region.
\citet{goyal2019skin} designed an automatic ensemble of DL methods, such as DeeplabV3+ \citep{everingham2010pascal} and Mask R-Convolutional Neural Network (R-CNN) \citep{he2017mask}, for generating the precise lesion boundaries. They combined the result of two models in three ways: a combination of both masks, picking the larger segmented area from the output of both methods, and picking a smaller area from those outputs. Finally, the authors discovered that the first method outperforms the other ensembling methods. 
\citet{al2018skin} proposed a Full Resolution Convolutional Network (FrCN) for the skin lesion segmentation, which learns full resolution features from each pixel of an input image. 
A segmentation method is realized by \citet{hawas2020oce} for the neutrosophic graph cut algorithm.
The initial clusters were obtained using Histogram-based Clustering Estimation (HBCE) with the corresponding centroids.
The genetic algorithm was applied to optimize the HBCE for getting the optimal threshold. 
Then, the Neutrosophic C-means (NCM) mapped the lesions into a Neutrosophic Set (NS) domain. Finally, for the lesion segmentation, the graph cut algorithm was a cost function.
\citet{amin2020integrated} segmented the lesion in two steps. In the first step, the authors performed preprocessing to resize the images to $240 \times 240 \times 3$ and convert the RGB into $Lab$ to select the luminance channel. Finally, in the second step, biorthogonal $2D$ wavelet transform and OTSU algorithm were applied for the lesion segmentation. 
\citet{xie2020mutual} produced a deep CNN, called mutual bootstrapping deep CNN (MB-DCNN). MB-DCNN has three networks, such as a coarse segmentation network (coarse-SN), a mask-guided classification network (mask-CN), and an enhanced segmentation network (enhanced-SN).
Coarse-SN was used to roughly segment the lesion and feed the label region to mask-CN to boost the classification task. 
\citet{al2020automatic} compared the performance of two variations of the UNet \citep{ronneberger2015u} model for the lesion segmentation, such as UNet without spatial dropout and UNet with spatial dropout.
In the end, the authors showed that augmentation and dropout, as regularization methods, with UNet, had less prone to overfitting and provided better-segmented lesion masks.
\citet{pour2020transform} offered a segmentation model based on CNN with CIELAB color space and transformed domain feature extraction. 
The authors initially implemented a scratch model inspired by UNet and FCN, then gradually improved the model by injecting features from the transformed domain and adding the input image color model CIElab.
They succeeded in coping with the constraints that included small data set, removal of artifacts, excessive data increase, and contrast stretching.
They also confirmed that the CNN model's performance with a domain transfer feature is better than the CNNs with a deep layer network.

\subsubsection{Methods for Lesion Recognition}
Many image analysis-based methods have already been proposed and developed by the researchers for dermoscopic image recognition, where the algorithms generally depend on the detection and extraction of low-level handcrafted features, such as colors, shapes, textures, and \textit{etc.} 
\citet{cheng2008skin} extracted different features from the first‐order histogram probability such as area, roundness (thinness), mean, standard deviation, skew, energy, and entropy. Finally, the authors applied different classifiers, namely quadratic discriminant analysis and Multilayer Perceptron (MLP), on the selected features by the Principal Components Analysis (PCA) \citep{mackiewicz1993principal}.
Different visual cues, such as ABCD (Asymmetry, Border, Color, and Differential structures) rule of dermoscopy \citep{nachbar1994abcd}, texture (neighboring gray-level dependence matrix, angular second-moment, and kinetics of skin lesions) were extracted by \citet{maglogiannis2009overview}. The authors also selected the features using the sequential backward floating selection, PCA, and generalized sequential feature selection algorithms. Finally, they employed MLP and Support Vector Machine (SVM) \citep{furey2000support} for the lesion classification.
\citet{oliveira2018computational} computed different local features employing the bag-of-features \citep{ghadiyaram2017perceptual} approach and texture features, where the authors selected the subset of features using a heuristic search approach. In the end, they applied SVM, Bayesian network, Decision Tree (DT) \citep{acharya2012automated}, and Artificial Neural Network (ANN) as classifiers.
\citet{hameed2020multi} developed an intellectual Multi-Class Multi-Level (MCML) classification algorithm employing two approaches, such as traditional machine learning and deep learning. In the former method, they applied preprocessing, segmentation, extraction of features, and classification. As a preprocessing, they removed the hair, black frames, and circle. Finally, the authors classified the texture and color features employing the ANN.
\citet{mporas2020color} applied a median filter followed by bottom-hat filtering to detect natural hair or similar to hair artifacts. They segmented the ROIs using the active contour model on the grayscale image. Finally, they extracted different color-based features for classification using the MLP and other Machine Learning (ML) algorithms. 

However, as described earlier, the lesion classification algorithms are very complex as they essentially rely on the handcrafted features extraction method, requiring prior knowledge \citep{fujisawa2019possibility} and lots of parameter tuning. 
Extensive feature engineering is the key to achieving better-performance from them, which is often impossible due to the presence of different artifacts in the dermoscopic images (see in Fig.~\ref{fig:artifacts}). 
The development of various CNN-based classifiers has achieved a remarkable result on the ImageNet dataset \citep{deng2009imagenet}.
Nowadays, in many computer vision problems, the contribution of both CNNs and DL techniques are undeniable \citep{guo2016deep}. CNN is an excellent feature extractor, which necessarily alleviates the manual feature engineering as in the algorithms mentioned above, therefore applying it to recognize medical images \citep{yadav2019deep}.
\citet{mahbod2019fusing} presented an ensemble-based model for CNNs that combines inter-and intra-architecture network fusion. The authors applied the fine-tuning of pre-trained VGGNet, AlexNet \citep{krizhevsky2012imagenet}, and two types of ResNet. Finally, the average prediction probability classification vectors from different sets were fused to provide the final prediction.
\citet{brinker2019enhanced} exercised ResNet-$50$ with transfer learning \citep{torrey2010transfer}. For the optimization of the model, they adopted three techniques. Firstly, they exclusively trained the adapted last layer, then fine-tuned all layers' parameters, and finally, a sudden increment of the learning rate at specific time steps during fine-tuning.
\citet{zhang2019attention} presented an Attention Residual Learning (ARL) CNN model for the skin lesion recognition, which was composed of multiple ARL blocks, a global average pooling, and a classification layer.
Each ARL block employed residual learning and novel attention learning mechanisms to improve its capability for discriminative representation. 
The authors proposed the attention learning mechanism, which aimed to utilize the intrinsic self-attention ability of DCNNs, i.e., using the feature maps learned from a high layer to generate a low-layer attention map instead of applying extra learnable layers.
An integrated framework for skin lesion boundary detection as well as for skin lesions classification was described by \citet{al2020multiple}. Firstly, a deep learning method, named FrCN, was used for the lesion boundary extraction. Then, geometric augmentation and transfer learning were integrated with four CNN networks, such as Inception-V3 \citep{szegedy2016rethinking}, ResNet-$50$, Inception-ResNet-V2, and DenseNet-201 \citep{huang2017densely} for the lesion classification. They also showed that segmented lesions improve lesion classification results. 
\citet{yilmaz2020benign} implemented three deep CNN models named  AlexNet, GoogLeNet, and ResNet-$50$. 
They compared classification performance as well as time complexity of the implemented models. 
For data augmentation, a  style-based Generative Adversarial Network (GAN) architecture was proposed by \citet{qin2020gan}. In the end, the authors applied ResNet-$50$, with transfer learning, for the lesion classification.
\citet{khan2020developed} proposed a model for the lesion classification, which included the localization of lesion ROI via faster region-based CNN, feature extraction, and feature selection by iteration-controlled Newton-Raphson method.
The ABC-based method was first used for contrast stretching and then used for lesion segmentation. 
DenseNet-$201$, via transfer learning, was used to extract deep-level features, and those features were classified using an MLP.
\citet{gessert2020skin} ensembled different DL methods, such as EfficientNets \citep{tan2019efficientnet}, SENet \citep{ hu2018squeeze}, and ResNeXt, by a selection strategy. They used multi-resolution input by multi-crop evaluation and two different cropping strategies. The encoding of metadata as a feature vector was concatenated with the dense (fully connected) neural network.
\citet{valle2020data} optimized the hyperparameter of two deep CNN models, ResNet-101-V2, and Inception-V4 employing transfer learning with data augmentation.
They select the best performing classifier using the ANOVA test \citep{scheffe1999analysis}.
Finally, the authors concluded that the transfer learning and ensembling model is a better choice for lesion classification.

\subsection{Our Contribution}
The above-discussions on the automatic skin lesion diagnosis methods confirm that the deep CNN approaches are commonly applied nowadays than the different systems relying on handcrafted features. The former approaches provide good reproducibility of results and boost diagnostic procedures' speed while being end-to-end methods.  
However, the CNN-based skin lesion analysis methods suffer from data scarcity to evade overfitting. The ensembling of different CNN architectures can mitigate those CNN's limitations, as proven by \citet{harangi2018skin}.    
In many articles \citep{  gessert2020skin, goyal2019skin,  ha2020identifying,harangi2018skin,  lee2018wonderm, mahbod2020transfer, pacheco2019skin,shahin2018deep}, the authors first trained different CNN models independently and then aggregated their outputs for developing ensembling models. Such an ensembling is tedious and time-consuming, leading to massive time and resources for training and testing. However, to eradicate those limitations, an end-to-end ensemble approach for skin lesion analysis without compromising state-of-the-art outcomes is highly essential. With the aforementioned thing in mind, this article aims to provide the following contributions:   

\begin{itemize}
    
    \item Develop an end-to-end ensembling model with dual encoders in our Dermo-DOCTOR framework, concatenating two different feature maps from those two encoders to broaden the lesion's depth information. Such a proposed network with two different encoders with the same input is likely to learn more discriminating features with limited training samples. 
    
    \item Incorporate segmented lesion ROIs for the recognition as ROIs enable the classifier to learn the abstract region and detailed structural description while avoiding surrounding healthy regions.
    
    \item Apply geometry- and intensity-based image augmentations and transfer learning to alleviate overfitting; the class rebalancing techniques to protect the classifier from being biased towards any particular class with more samples.

    \item Develop and compare two other networks for detection (UNet and FCN8s) and recognition (ResNet-$50$ and Xception \citep{chollet2017xception}) under the same experimental settings. 
    
    \item Demonstrate state-of-the-art lesion detection and recognition results, to our best knowledge, on two IEEE International Symposium on Biomedical Imaging (ISBI) datasets, such as ISIC-$2016$ and ISIC-$2017$, having a different number of classes. 
    
    \item Implement a possible application of our Dermo-DOCTOR, deploying its trained weights, which runs in a web browser (see in YouTube\footnote{Dermo-DOCTOR App: \url{https://bit.ly/Dermo-DOCTOR}}). 

\end{itemize}

The rest of the paper is structured accordingly. We explain the design of the Dermo-DOCTOR framework and datasets in section \ref{materialandMethods}. The results and discussions of the extensive experiments, with the proper interpretation, are reported in section \ref{ResultsandDiscussion}. Finally, we conclude the article in section \ref{Conclusion}.

\section{Materials and Methods}
\label{materialandMethods}
This section manifests the materials and methods, describing the proposed Dermo-DOCTOR pipeline in Section \ref{Framework}. We explain the utilized datasets, integral preprocessing, and the proposed network in Subsections \ref{Datasets}, \ref{Preprocessing}, and \ref{Concurrent}, respectively.  Sections \ref{app} and \ref{Training} respectively describe our web application and network's training protocol. 

\subsection{Proposed Framework}
\label{Framework}
\begin{figure*}[!ht]
  \centering
  \subfloat{\includegraphics[width=15cm, height= 4.8cm]{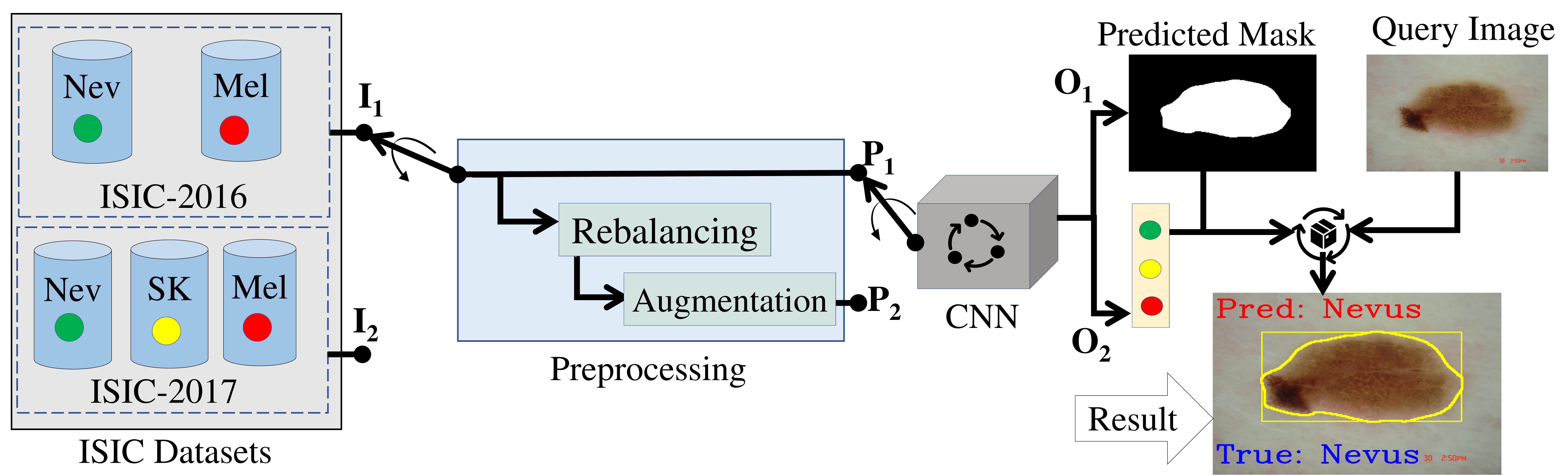}}
  \caption{The proposed pipeline for concurrent detection and recognition systems, where the preprocessing has incorporated with the proposed network to build a precise diagnostic system. An input $I_1$ or $I_2$ generates two outputs $O_1$ and $O_2$, where $O_1$ and $O_2$ respectively denote the segmentation and recognition results.}
  \label{fig:framework}
\end{figure*}

The overall Dermo-DOCTOR framework is illustrated in Fig.~\ref{fig:framework}.
We utilize two different input types (either $I_1$ or $I_2$), where $I_1$ or $I_2$ is a binary or a multi-class categorization task. 
An input, either $I_1$ or $I_2$, generates two different outputs, such as segmentation ($O_1$) and recognition ($O_2$). 
The outputs $O_1$ and $O_2$ are then processed to provide lesion detection and recognition results. We process the predicted lesion masks to generate the bounding box around the lesion, naming lesion detection.
However, different crucial integral parts of the Dermo-DOCTOR are explained in the following subsections.

\subsubsection{Datasets}
\label{Datasets}
\begin{figure*}[!ht]
  \centering
  \subfloat[]{\includegraphics[width=6cm, height= 4.8cm]{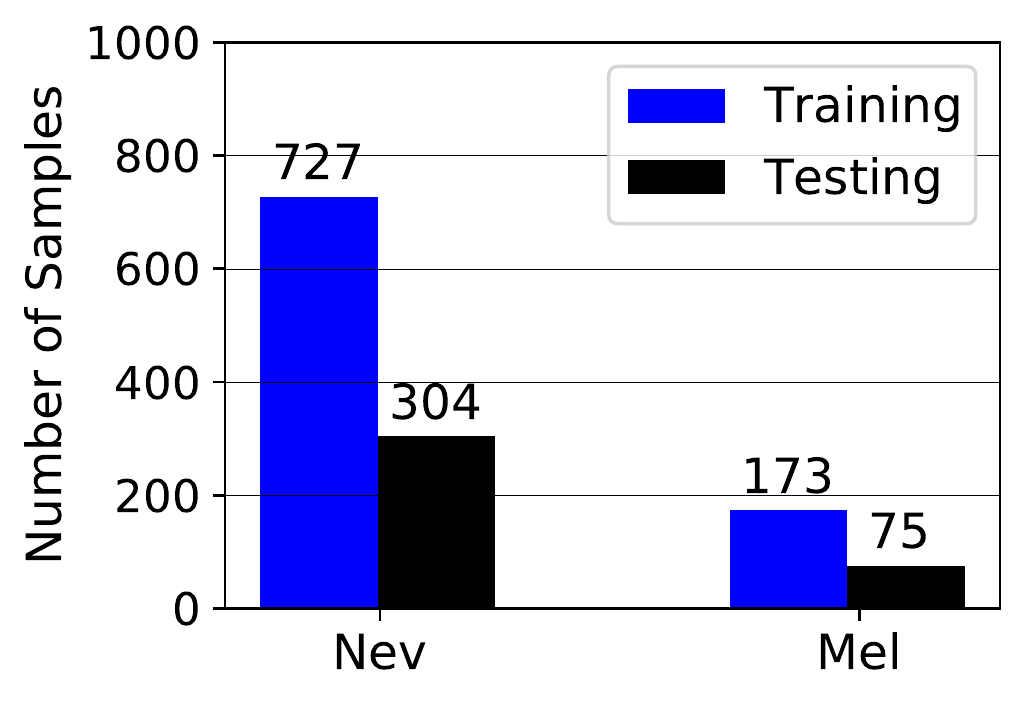}} \hspace{0.25cm}
  \subfloat[]{\includegraphics[width=10cm, height= 4.8cm]{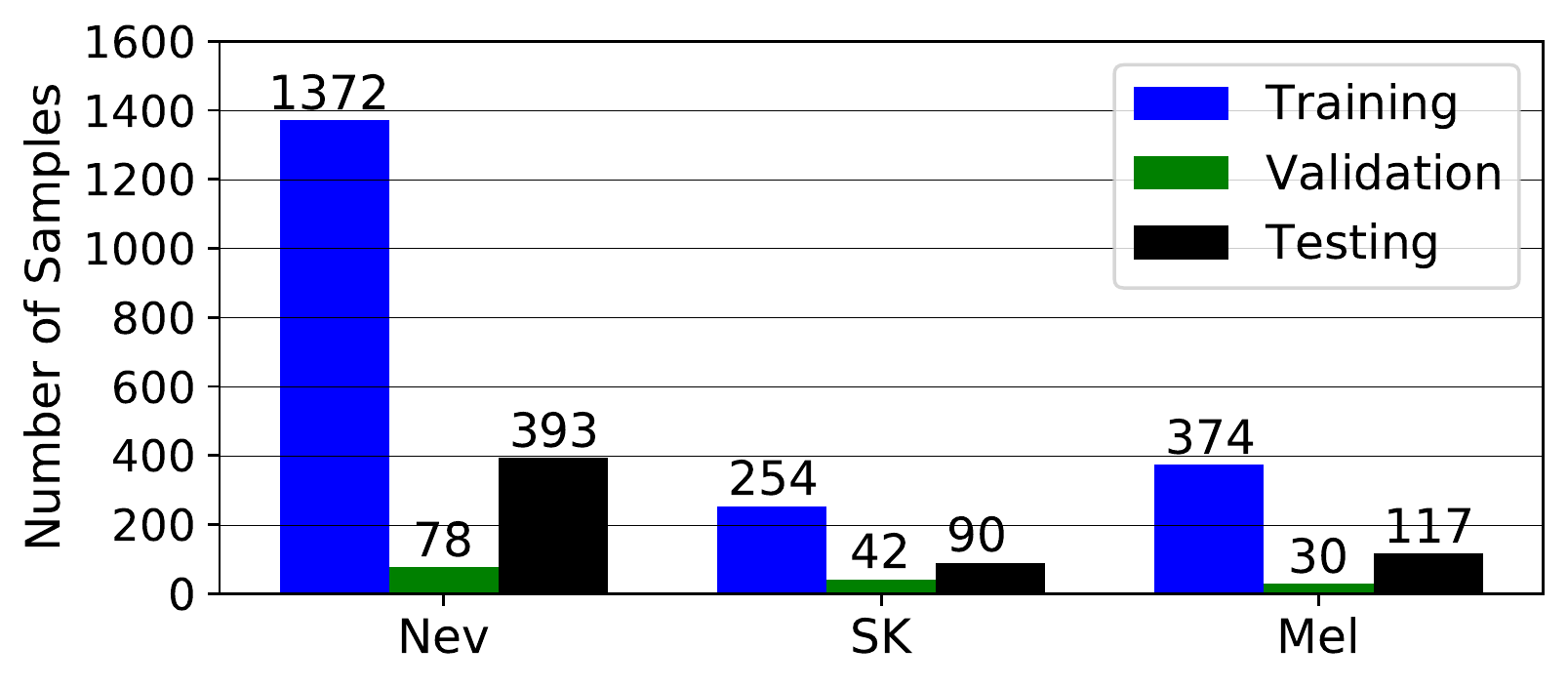}}
  \caption{The distributions of the utilized ISIC datasets, where (a) is for ISIC-2016 and (b) is for ISIC-2017. The validation set for the ISIC-2016 is not available publicly (see in the left figure). }
  \label{fig:datasets}
\end{figure*}

Two different datasets, such as ISIC-2016 \citep{gutman2016skin} and ISIC-2017 \citep{codella2018skin}, are used to validate our proposed pipeline, whose class-wise distributions are presented in Fig.~\ref{fig:datasets}. 
The ISIC-2016 contains a binary class, aiming to classify as either Nevus (Nev) or Melanoma (Mel), explicating that class samples are imbalanced ($4.2:1$ for $Nev: Mel$).
On the other hand, the ISIC-2017 is a multi-class categorization task, intending to classify as either Nevus (Nev), or Seborrheic Keratosis (SK), or Melanoma (Mel). 
The distribution of ISIC-2017 also tells that class samples are highly imbalanced, where $Nev: SK: Mel$ in the training set is $11.8:2.8:1$. 
The training samples' imbalanced distribution makes the classifier biased towards the particular class with more samples, mitigated in the proposed pipeline by adopting two techniques (see in subsection \ref{Preprocessing}).

\subsubsection{Preprocessing}
\label{Preprocessing}
We have applied class rebalancing and different image augmentations (both geometry- and intensity-based) as a preprocessing, which are concisely explained as follows:

\textbf{Rebalancing.}
The class imbalance is a common phenomenon in the medical imaging domain as manually annotated images are very complex and arduous to achieve \citep{harangi2018skin}.
Such a class imbalance can be partially overcome using two commonly used approaches, such as the data-level method and algorithmic level method \citep{he2009learning}. 
We have combined additional images to the underrepresented class from the ISIC archive \citep{ISICArch3} and weighted the loss function. For weighing the loss function, we apply $W_i = N_i/N$, where $W_i$, $N$, and $N_i$ are the weight for $i^{th}$ class, the total sample numbers, and the sample numbers in the $i^{th}$ class, respectively.

\textbf{Augmentation.}
One of the crucial challenges in the medical imaging domain is coping with the small datasets, such as in the ISIC datasets \citep{harangi2018skin}. However, we have applied different augmentations based on geometric transformations, such as rotation, flipping, shifting, zooming,  and image processing functions, such as gamma, logarithmic, sigmoid corrections, and stretching, or shrinking the intensity levels.

However, the input $I_1$ or $I_2$ produces the lesion recognition output by applying two preprocessing types: the $P_1:$ only segmentation and the $P_2:$ rebalancing and augmentation with segmentation.

\subsubsection{Proposed Network}
\label{Concurrent}
\begin{figure*}[!ht]
  \centering
  \subfloat{\includegraphics[width=16.4cm, height= 13.1cm]{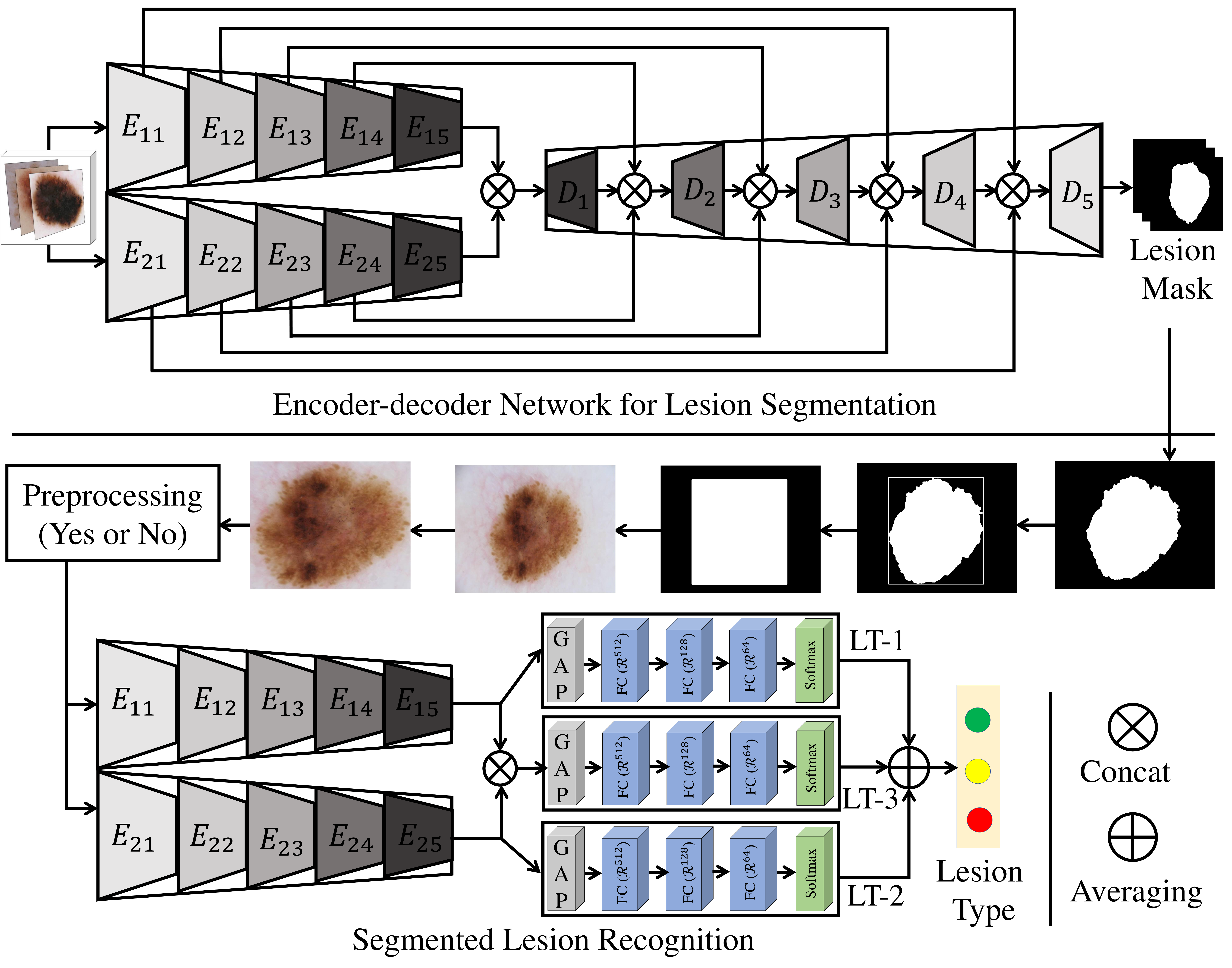}} 
  \caption{The proposed network for the Dermo-DOCTOR application, where the first encoder-decoder sub-network is applied for the lesion segmentation, while the second sub-network is employed for the lesion recognition. The segmented lesion masks are used for ROI extraction for further classification and detection utilizing the bounding boxes around the lesions.}
  \label{fig:network}
\end{figure*}

Nowadays, CNN-based methods outperform the radiologists with high values of balanced accuracy as proven in \citep{kermany2018identifying,rajpurkar2017chexnet}.
Nevertheless, they trained the models with an enormous number of annotated images.
However, CNNs may be obliquely limited when employed with highly variable and distinctive image datasets with limited samples and having inter-class homogeneity and intra-class heterogeneity, as in dermoscopic ISIC datasets \citep{codella2018skin,gutman2016skin}. 
Ensembling the network is likely to alleviate data scarcity limitation for the CNN training \citep{dolz2020deep,harangi2018skin,kumar2016ensemble,moitra2020prediction,savelli2020multi}. 
In this context, we propose a CNN-based end-to-end ensemble network for simultaneous lesion detection and recognition, consisting of two encoders, a decoder, and three Fully-connected Layers (FCLs), as shown in Fig.~\ref{fig:network}. The different parts of the proposed dual encoder network are explained in the following paragraphs.  

\begin{figure*}[!ht]
  \centering
  \subfloat{\includegraphics[width=16cm, height= 10cm]{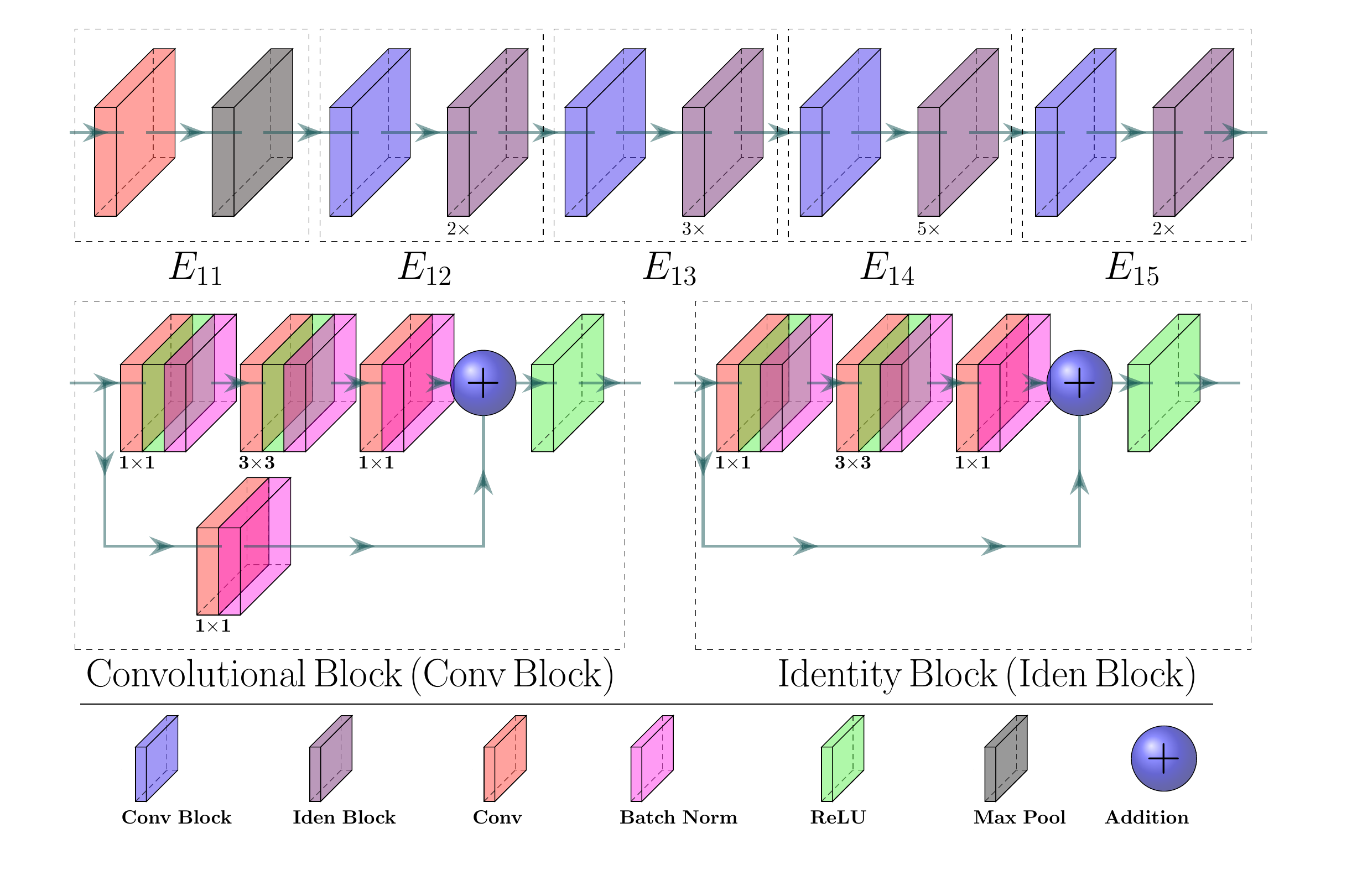}}
  \caption{The encoder-1 of the proposed network, where the Conv and Iden blocks are stacked on top of each other. The notation ($n\times $) under the Iden block denotes the number of repetitions ($n$-times).}
  \label{fig:encoder1}
\end{figure*}

\textbf{Encoder-1.} 
The first encoder ($f^{en-1}$) in the proposed network is presented in Fig.~\ref{fig:encoder1}.   
It includes Identity (Iden) and Convolutional (Conv) blocks, applying the skip connections \citep{he2016deep} in both blocks. 
There are two main advantages of such a skip connection. Firstly, the lack of regularization of the new layers does not affect their performance, and secondly, the new layers are not nil even when they are regulated.
In encoder-1, an input convolution is adopted before Iden and Conv blocks, followed by a max-pooling.
By stacking these blocks on top of each other, encoder-1 has been designed for getting a lesion feature map (see in Fig.~\ref{fig:encoder1}). 
The output feature map of the encoder-$1$ is defined as $X_{en-1}=f^{en-1}(I_{in})$, where $X_{en-1} \in \mathcal{R}^{B\times H \times W \times D}$, and $B$, $H$, $W$, $D$, and $I_{in}$ respectively denote the batch size, height, width, depth (channel), and input batch of images.  
The encoder-$1$ is divided into five sub-blocks ($E_{1n}$ and $n=1, 2, ..., 5$), while the input image resolutions in each sub-block are down-sampled in half of the input resolutions.
In the segmentation sub-network, each sub-block's outputs ($E_{1n}$ and $n=1, 2, ..., 5)$ will be used as an input for the skip connections to regain the lost spatial information due to pooling in the encoders.

\begin{figure*}[!ht]
  \centering
  \subfloat{\includegraphics[width=16cm, height= 7cm]{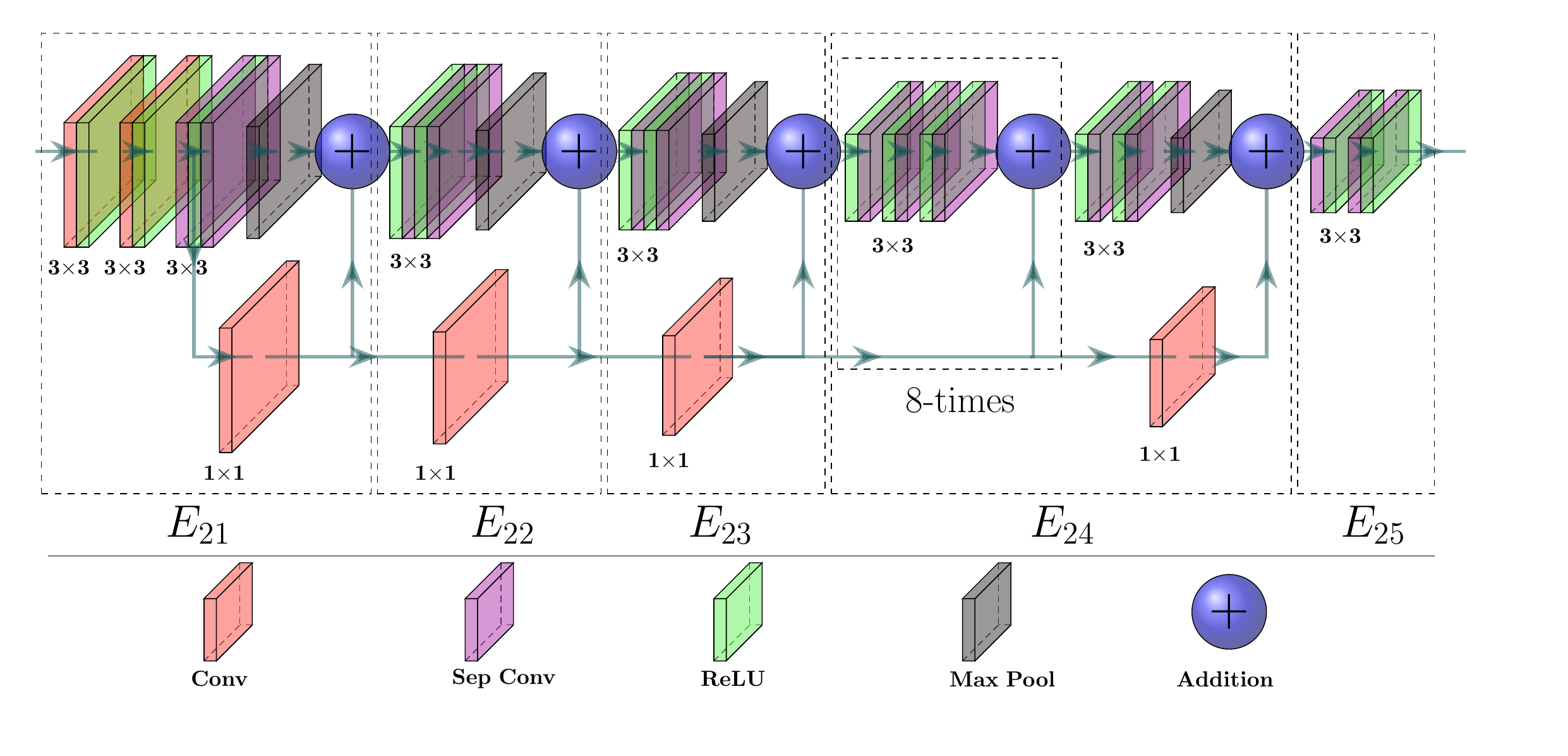}} 
  \caption{The encoder-2 of the proposed network, where depth-wise separable convolutions \citep{chollet2017xception} were employed instead of traditional convolutions to make it lightweight for real-time applications.}
  \label{fig:encoder2}
\end{figure*}

\textbf{Encoder-2.}
Within the encoder-2 ($f^{en-2}$), three block components are employed, such as entry flow, middle flow, and exit flow \citep{chollet2017xception}. Fig.~\ref{fig:encoder2} depicts the constructional details of the encoder-2.
The batch of input images firstly passes through the input flow, then the central flow, repeated eight times ($8\times $), and finally through the exit flow.
All flows employ Depth-wise Separable Convolution (DwSC) \citep{chollet2017xception} and residual connections. 
The former has been used to create a lightweight network, while the latter has the advantages discussed earlier in encoder-1.
The output feature map of the encoder-2 is defined as $X_{en-2}=f^{en-2}(I_{in})$, where $X_{en-2} \in \mathcal{R}^{B\times H \times W \times D}$, and $B$, $H$, $W$, $D$, and $I_{in}$ respectively denotes batch size, height, width, depth (channel), and input batch of images.
The encoder-2 is also divided into five sub-blocks ($E_{2n}$ and $n=1, 2, ..., 5$), in which input resolutions are also down-sampled into half the resolutions of each sub-block. 
Every sub-blocks ($E_{2n}$ and $n=1, 2, ..., 5$) are then used as the skip connections, when it is decoded in the detection sub-network.

\begin{figure*}[!ht]
  \centering
  \subfloat{\includegraphics[width=16cm, height= 6.5cm]{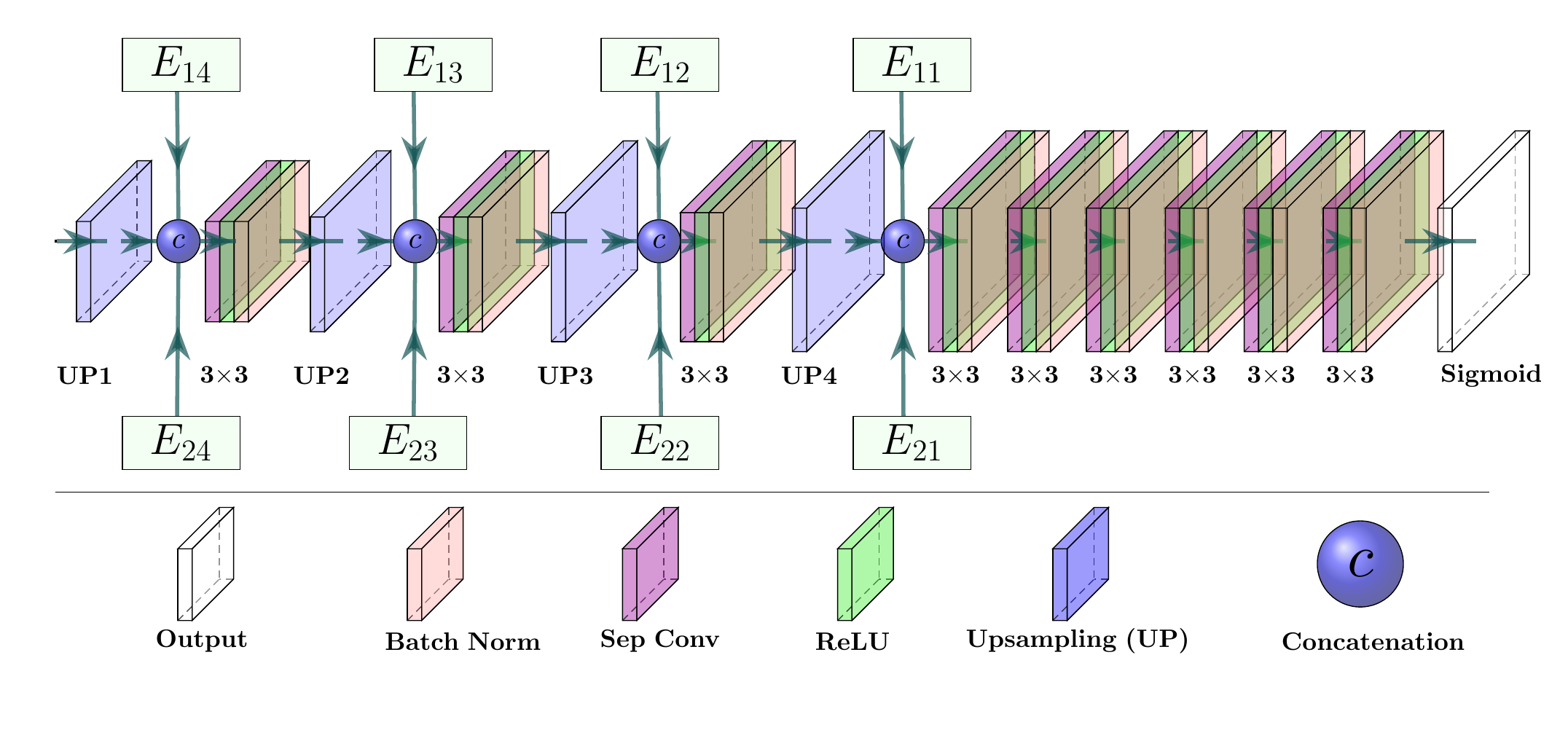}} 
  \caption{The decoder of the proposed detection sub-network for reconstructing a segmentation mask with the input resolution from the encoders' low-resolution features.}
  \label{fig:decoder}
\end{figure*}

\textbf{Detection Sub-Network.} 
The decoder semantically projects the salient features of lower resolution from the encoders onto the pixel space having a higher resolution to achieve a semantic lesion pixel label \citep{garcia2018survey,long2015fully,ronneberger2015u}.
The reduced feature maps (to attain spatial invariance) from the encoder often cause a loss in spatial resolution, bringing zigzag edge information, coarseness, checkerboard artifacts, and over- and under-segmentation in the segmented masks \citep{hasan2021drnet,hasan2020dsnet,long2015fully, odena2016deconvolution,ronneberger2015u}.
Although there are many approaches to alleviate these problems in the segmentation \citep{al2018skin,badrinarayanan2017segnet, hasan2021drnet,hasan2020dsnet, long2015fully,odena2016deconvolution, ronneberger2015u}, there is still room for performance improvement. 
In our detection sub-network, the obtained outputs from two encoders are concatenated channel-wise for enlarging the depth representation of the feature map, which is named as a Fused Feature Map (FFM), where $FFM \in \mathcal{R}^{B\times H \times W \times 2D}$.  
We have applied skip connections, inspired by the UNet, to tackle the subsampling limitations. The $FFM \in \mathcal{R}^{B\times H \times W \times 2D}$ is an input to the decoder of our detection sub-network (see in Fig.~\ref{fig:decoder}). 
Unlike the earlier networks, we skip the features from two different encoders to recover the lost spatial information (see in Fig.~\ref{fig:decoder}). 
The channel-concatenation in each stage of decoder is presented as $[E_{1n} \concat E_{2n} \concat D_{n} ]$, where $E_{1n}$, $E_{2n}$, $D_{n}$, and $\concat$ respectively denote skipped feature maps from encoder-1 and encoder-2, decoder feature map (at $n^{ht}$ stage), and channel concatenation. $E_{1n}$, $E_{2n}$, and $D_{n}$ are the same scaled feature maps and $n=4, 3, ..., 1$. 
Such a dual encoder skipped feature has enhanced depth information, which is likely to improve the segmentation accuracy by better retrieving the lost spatial information.   
Besides, we employ \textit{batch normalization} \citep{Sergey} to overcome the internal covariate shift in the training phase. We also compact our network's design, employing a DwSC \citep{chollet2017xception} in place of standard convolution. 
We decrease the parameters by a factor of $(1/N + 1/K^2)$ for each convolution in our Dermo-DOCTOR, where $N$ and $K$ respectively indicate the filter number and kernel size \citep{hasan2021detection}.

\textbf{Recognition Sub-Network.}
Different feature maps from the encoder-1, encoder-2, and FFM are classified into desired categories applying the FCLs.
We employ a Global Average Pooling (GAP) layer \citep{lin2013network} before the FCL for vectorizing the 2D feature maps into a single long continuous linear vector, as it improves generalization and prevents overfitting \citep{lin2013network}.  
Additionally, each FCL layer is followed by a \textit{dropout} layer \citep{srivastava2014dropout} as a regulariser, where we randomly set $50.0\,\%$ neurons of the FC layer to zero during the training. 
However, the two different feature maps of two different encoders are utilized to recognize the Lesion Type (LT) separately using the two FCLs, termed as LT-1 and LT-2 (see in Fig.~\ref{fig:network}). 
Besides, the FFM also generates LT-3, as explicated in Fig.~\ref{fig:network}. 
Finally, the output probability ($O_{j=1,2}$) is the average of the LT-1, LT-2, and LT-3.
The output ($O_{j=1,2}$) lies in $N$-dimensional space, where $O_1 \in \mathcal{R}^{N=2}$ and $O_2 \in \mathcal{R}^{N=3}$ respectively for the inputs $I_1$ or $I_2$ by applying the proposed preprocessing (either $P_1$ or $P_2$). It is noteworthy that the output lesion class ($O_{j=1,2}$) is obtained from the end-to-end training.

\subsection{Designing of Web Application} 
\label{app}
\begin{figure*}[!ht]
  \centering
  \subfloat{\includegraphics[width=16cm, height= 8cm]{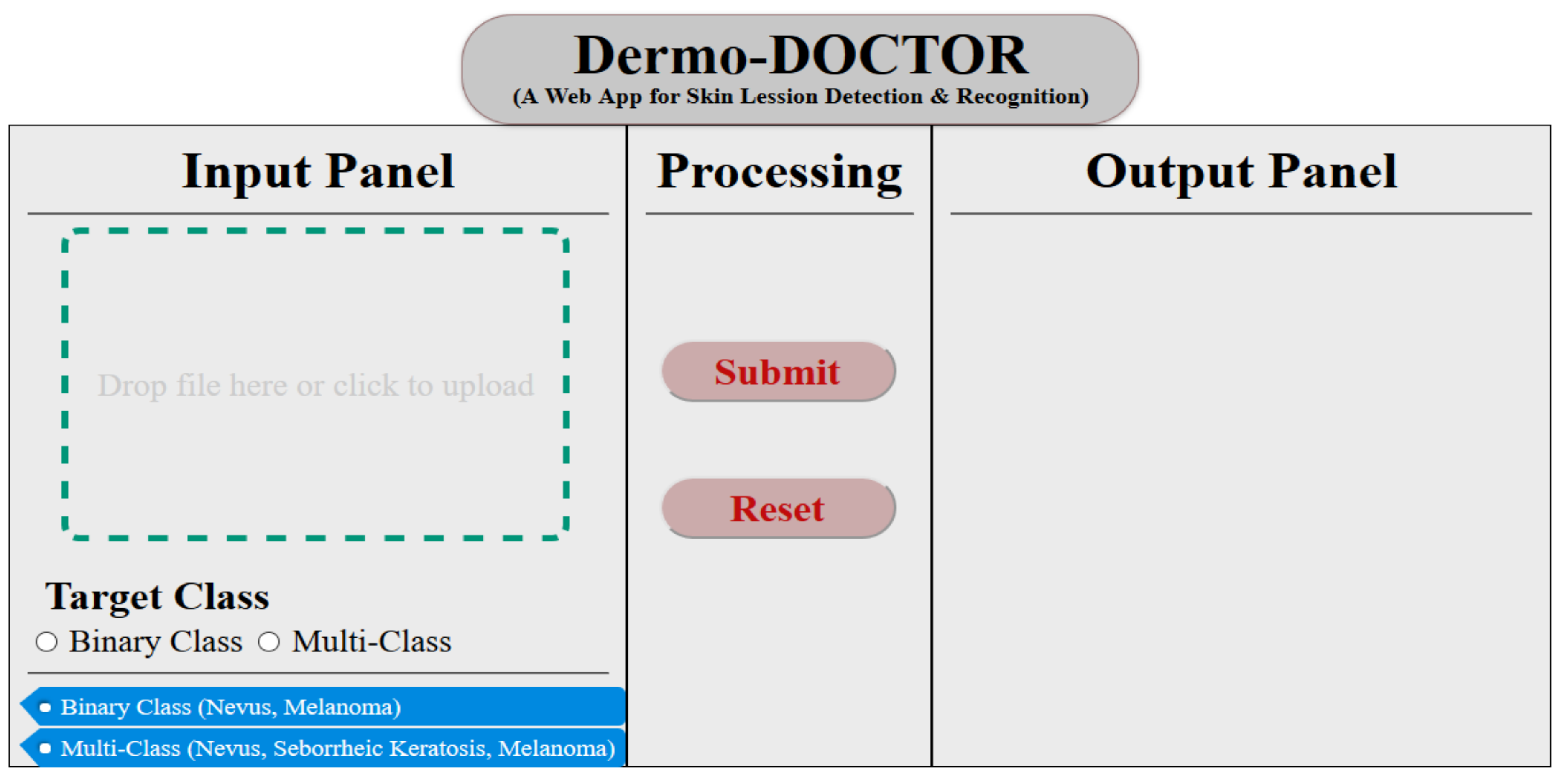}} 
  \caption{The dermo-DOCTOR prototype, where the user can select or drag a dermoscopic image (png, jpg, bmp, or jpeg) as an input. The desired number of output classes also can be selected in the input panel. The processing and output panels are dedicated to select the process types and display the recognized class with the probabilities.}
  \label{fig:appDermo}
\end{figure*}

The proposed web application named Dermo-DOCTOR, for the end-users, is depicted in Fig.~\ref{fig:appDermo}. 
We utilize browser-supported languages such as Hypertext Markup Language (HTML), Cascading Style Sheets (CSS), and Javascript, \textit{etc.} for developing the  Dermo-DOCTOR application. 
A python web framework package, called Flask, is used for developing an application by deploying our proposed CNN-based detection and recognition models and their trained weights. We apply HTML and CSS to design a graphical user interface with three panels, such as an input panel, a processing panel, and an output panel (see in Fig.~\ref{fig:appDermo}). The user can select the query image (drag-and-drop or direct upload) and the number of query classes (both binary and multi-class) in the input panel. Then, the user can also start the process or reset the selections in the processing panel. The return results from the host machine are displayed in the output panel.

\subsection{Training Protocol}
\label{Training}
The encoder kernels are initialized with the pre-trained ImageNet weights, whereas the decoder kernels are initialized with the \say{he normal} distribution \citep{henormal}.  
The Aspect Ratio (AS) distribution tells that most of the images in ISIC-2016 and ISIC-2017 datasets have an AS of $3:4$. Therefore, we resize all the images to $192 \times 256$ pixels using the nearest-neighbor interpolation for the detection.
Again, the AS distribution of both datasets' extracted lesion ROIs reveals that most of the ROIs have an AS of $1:1$. 
Hence, we again resize the lesion ROIs to $192 \times 192$ pixels using a nearest-neighbor interpolation for the recognition. 
Additionally, we have standardized and rescaled the training images to $[0 \, 1]$ for both detection and recognition. 
We employ Eq.~\ref{eq:DSNETLoss} as a loss function and intersection over union as a metric for training the detection sub-network of the proposed Dermo-DOCTOR. 
\begin{equation} \label{eq:DSNETLoss}
L\,(y,\hat{y})= 1-\frac{\displaystyle\sum_{i=1}^{N}y_i\times \hat{y_i}}{\displaystyle\sum_{i=1}^{N}y_i+\displaystyle\sum_{i=1}^{N}\hat{y_i}-\sum_{i=1}^{N}y_i\times \hat{y_i}}-\frac{1}{N} \sum_{i=1}^{N}[y_i \log \hat{y_i} + (1-y_i)\log (1-\hat{y_i}) ],
\end{equation}
\noindent where $y$ and $\hat{y}$, $N$ respectively denote the true and predicted label, the total pixel numbers.  
In Eq.~\ref{eq:DSNETLoss}, $\log\hat{y_i}$ and $\log(1-\hat{y_i})$ are the estimation of log-likelihood of pixel being lesion or not, respectively. The product of $y$ and $\hat{y}$ in Eq.~\ref{eq:DSNETLoss} is the estimation of similarity (intersection) between true and predicted lesion masks. We employ categorical cross-entropy as a loss function and accuracy as a metric for training the recognition sub-network.

\section{Results and Discussion}
\label{ResultsandDiscussion}
This section bestows different lesion detection results and subsequent recognition in Subsections \ref{segmentation} and \ref{classification}, respectively. The segmented lesion masks are utilized for ROI extraction to classify and detect the bounding boxes around the lesions.

\subsection{Results for Detection}
\label{segmentation}
Firstly, we exhibit the quantitative and qualitative segmentation results, applying the proposed Dermo-DOCTOR and two other well-known networks: the UNet and the FCN8s. Secondly, we compare our outcomes with several state-of-the-art results utilizing the same datasets. To quantify the segmentation correctness, we use mean Recall (mRc), mean Specificity (mSp), and mean Intersection over Union (mIoU), which are defined in Eq.~\ref{eq:metrics}.
\begin{equation} 
\label{eq:metrics}
\begin{split}
mRc= \frac{1}{M\times N}\displaystyle\sum_{i=1}^{N} \displaystyle\sum_{j=1}^{M}\frac{TP_{ij}}{TP_{ij}+FN_{ij}},  \\
mSp= \frac{1}{M\times N}\displaystyle\sum_{i=1}^{N} \displaystyle\sum_{j=1}^{M}\frac{TN_{ij}}{TN_{ij}+FP_{ij}},  \\
mIoU= \frac{1}{M\times N}\displaystyle\sum_{i=1}^{N} \displaystyle\sum_{j=1}^{M}\frac{TP_{ij}}{TP_{ij}+FN_{ij}+ FP_{ij}}, 
\end{split}
\end{equation}
where $M$ and $N$ denote the pixel and sample numbers, whereas $TP$, $TN$, $FN$, and $FP$ indicate true positive (lesion as a lesion), true negative (background as a background), false negative (lesion as a background), and false positive (background as a lesion), respectively.

Table~\ref{tab:segResults} confers the segmentation results of three methods on two separate datasets: the ISIC-2016 and the ISIC-2017.  
\begin{table*}[!ht]
\centering
\caption{Quantitative segmentation results on ISIC-$2016$ and ISIC-$2017$ test datasets using the Dermo-DOCTOR, UNet, and FCN8s. The winner metrics for the ISIC-2016 are presented in bold font, whereas they are underlined for the ISIC-2017.}
\begin{tabular}{ccccc}
\hline
\rowcolor[HTML]{C0C0C0} 
\cellcolor[HTML]{C0C0C0}                                                                                & \cellcolor[HTML]{C0C0C0}                                                                             & \multicolumn{3}{c}{\cellcolor[HTML]{C0C0C0}Performance metrics} \\ \cline{3-5} 
\rowcolor[HTML]{C0C0C0} 
\multirow{-2}{*}{\cellcolor[HTML]{C0C0C0}\begin{tabular}[c]{@{}l@{}} Testing\\ datasets  \end{tabular}} & \multirow{-2}{*}{\cellcolor[HTML]{C0C0C0}\begin{tabular}[c]{@{}l@{}}Models\end{tabular}} & mRc                 & mSp                 & mIoU                \\ \hline
                &                                                                                   Dermo-DOCTOR          &         $\mathbf{0.92\pm0.13}$            &  $0.97\pm0.07$                   &    $\mathbf{0.85\pm0.12}$                 \\ \cline{2-5}
                                                                              & UNet                                                                                           &         $0.86\pm0.13$            &  $\mathbf{0.98\pm0.04}$                   &    $0.83\pm0.13$                 \\ \cline{2-5}

\multirow{-3}{*}{\begin{tabular}[c]{@{}l@{}}ISIC-$2016$\end{tabular}}   & FCN8s                                                                                           &         $0.84\pm0.15$            &  $0.98\pm0.05$  &    $0.80\pm0.14$                        \\ \hline \hline

  & Dermo-DOCTOR  & $\underline{0.86\pm0.17}$ & $\underline{0.97\pm0.07}$& $\underline{0.80\pm0.17}$  \\ \cline{2-5} 
                          
                        &  UNet   & $0.83\pm0.21$ & $0.97\pm0.07$& $0.76\pm0.21$   \\ \cline{2-5} 
\multirow{-3}{*}{\begin{tabular}[c]{@{}l@{}}ISIC-$2017$\end{tabular}}                        &  FCN8s   & $0.86\pm0.18$ & $0.93\pm0.10$& $0.71\pm0.17$  \\ \hline
 
\end{tabular}
\label{tab:segResults}
\end{table*}
The mean overlapping between the actual and predicted masks are as high as $85.0\,\%$ and $80.0\,\%$ for the ISIC-2016 and ISIC-2017. 
The Dermo-DOCTOR mIoU beats nearby UNet by the margins of $2.0\,\%$ and $4.0\,\%$ accordingly for the ISIC-2016 and ISIC-2017 datasets.
The mean type-I and type-II errors for ISIC-2016 of the Dermo-DOCTOR are $3.0\,\%$ and $8.0\,\%$, respectively, outperforming the UNet by a margin of $6.0\,\%$ concerning the type-II error, while defeated by a margin of $1.0\,\%$ for type-I error. 
However, $1.0\,\%$ error in mSp is quite acceptable in segmentation as the lesion edge information will be preserved with less over-segmentation. 
Similarly, for ISIC-2017, the Dermo-DOCTOR exceeds UNet by a border of $3.0\,\%$ for type-II error, while they have similar type-I errors. 
For both the datasets, the FCN8s is a defeating method as it produces low metrics than the other two methods (see in Table~\ref{tab:segResults}).
In contrast, the proposed Dermo-DOCTOR provides better-segmented masks, leading to better lesion detection and recognition.

\begin{figure*}[!ht]  
\centering 
\includegraphics[width=15.5cm,height=12cm]{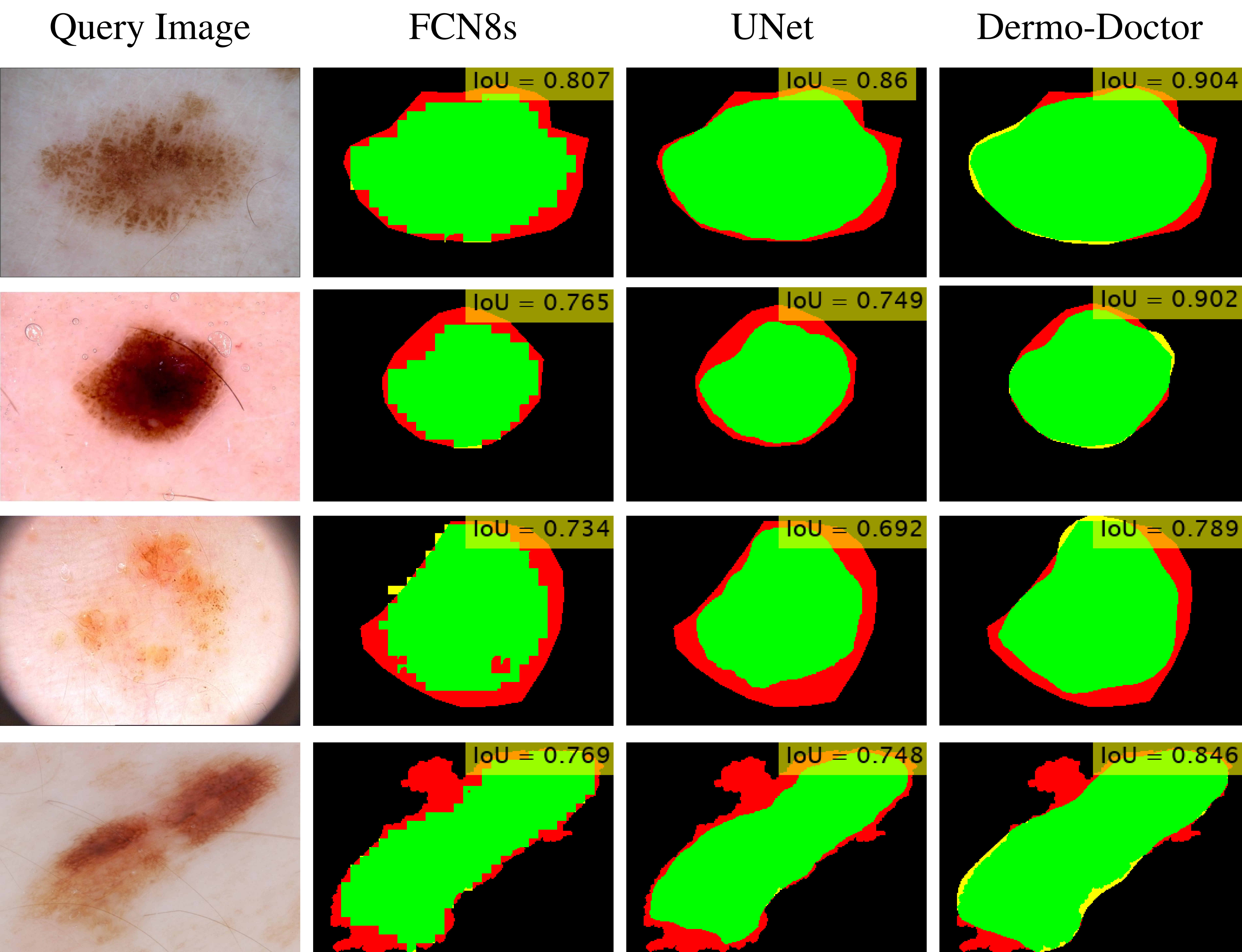}   
\caption {Qualitative segmentation results using the FCN8s, UNet, and Dermo-DOCTOR networks. Green, Red, and Yellow colors indicate the TP, FN, and FP regions, respectively. The top-right IoUs are given for quantitative evaluation.}
\label{fig:qualimodelcompare} 
\end{figure*}

\begin{figure*}[!ht]  
\centering  
\includegraphics[width=16cm,height=12cm]{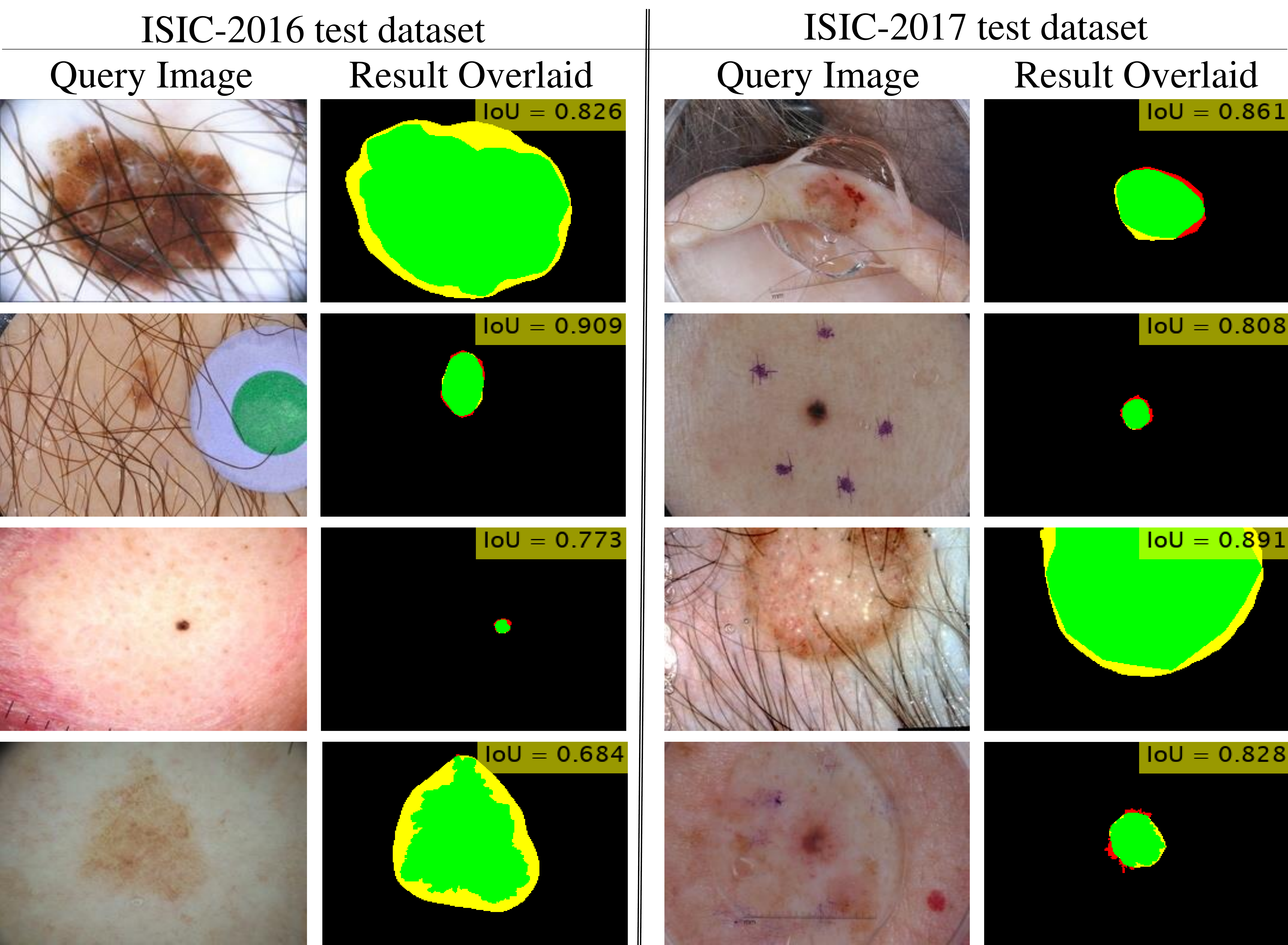}   
\caption{Qualitative segmentation results on two different test datasets employing our Dermo-DOCTOR, where the TP, FN, and FP regions are respectively denoted by the Green, Red, and Yellow colors. The top-right IoU exhibits quantitative validation.}
\label{fig:quali2}
\end{figure*}

Fig.~\ref{fig:qualimodelcompare} depicts several qualitative results from the proposed Dermo-DOCTOR, UNet, and FCN8s. 
The segmented lesions from FCN8s suffer from checkerboard artifact, providing zigzag lesion boundaries (see in Fig.~\ref{fig:qualimodelcompare}). 
The possible cause of such poor results is that the low-resolution features across the high-resolution feature map are not constant due to the non-divisible upscaling factor, generating checkerboard artifacts in the lesion masks \cite{hasan2021drnet,hasan2020dsnet}.
However, the lesion masks from the UNet and Dermo-DOCTOR are blessed with smoother lesion boundaries,  which are the image-based classifiers' salient features.
The quantitative and qualitative results of the UNet point that it undergoes from the additional false-negative region, leading to under-segmentation. 
However, the proposed Dermo-DOCTOR achieves lower false-positive and false-negative regions relative to the other implementations. 
As in the proposed Dermo-DOCTOR, the dual encoders can learn more salient features compared to the single encoder in UNet and FCN8s, especially when training with a small dataset.
Again, Fig.~\ref{fig:quali2} illustrates the segmentation results of several challenging images on the ISIC-2016 and ISIC-2017 datasets. 
The lesion segmentation results in Fig.~\ref{fig:quali2} describe that the segmented lesion masks are precise even the query images contain different artifacts (see in Fig.~\ref{fig:artifacts}), and even the ROIs are tiny in size. 
Although the extracted ROIs have few false-positive regions (yellow color) with negligible false-negative regions (red color), it is not surprising for further lesion classification, as they preserve the lesion boundaries. 

Table~\ref{tab:compareSEG} exhibits lesion segmentation results from the proposed Dermo-DOCTOR and other state-of-the-art methods, trained and tested on the same ISIC datasets.  
\begin{table*}[!ht]
\footnotesize
\caption {Comparative lesion segmentation results for the proposed Dermo-DOCTOR and other state-of-the-art methods on both the ISIC-2016 and ISIC-2017 test datasets.}
  \begin{tabular}{ l | c c c | c  c c c}
        \hline
        \rowcolor[HTML]{C0C0C0} 
        \cellcolor[HTML]{C0C0C0}{\color[HTML]{000000} } & \multicolumn{3}{c|}{\cellcolor[HTML]{C0C0C0}ISIC-2016 test dataset} & \multicolumn{3}{c}{\cellcolor[HTML]{C0C0C0}ISIC-2017 test dataset} \\ \cline{2-7} 
        \rowcolor[HTML]{C0C0C0} 
        \multirow{-2}{*}{\cellcolor[HTML]{C0C0C0}{\color[HTML]{000000} Segmentation Methods}} & mIoU & mRc & mSp & mIoU & mRc & mSp \\ \hline
        FCN ensemble \citep{yuan2017automatic} & $0.84$  & $0.91$ & $0.96$  & - & - & -  \\ \hline
        Fusion Structure \citep{tang2019multi} & $0.85$  & $0.92$ & $0.96$  & - & - & -  \\ \hline
        DCL-PSI \citep{bi2019step} & $0.85$ & $\mathbf{0.93}$ & $0.96$  & $0.72$ & $0.80$ & $0.94$  \\ \hline

        DSNet \citep{hasan2020dsnet} & - & - & -  & $0.77$ & $\mathbf{0.87}$ & $0.95$  \\ \hline
        
        HRFB \citep{xie2020skin} & $0.85$ & $0.87$ & $0.96$  & $0.78$ & $\mathbf{0.87}$ & $0.96$  \\ \hline
        
        iFCN \citep{ozturk2020skin} & - & - & -  & $0.78$ & $0.85$ & $\mathbf{0.98}$  \\ \hline
        
        \textbf{Proposed Dermo-DOCTOR (2020)}    &  $\mathbf{0.85}$     & $\underline{0.92}$    & $\mathbf{0.97}$  &  $\mathbf{0.80}$     & $\underline{0.86}$    & $\underline{0.97}$ \\ \hline
\multicolumn{7}{l}{\scriptsize{\textbf{DCL-PSI:} Deep Class-specific Learning with Probability based Step-wise Integration } } \\
\multicolumn{7}{l}{\scriptsize{\textbf{HRFB:} High-Resolution Feature Block } } \\
\multicolumn{7}{l}{\scriptsize{\textbf{iFCN:} improved Fully Convolutional Network} } \\
  \end{tabular}
\label{tab:compareSEG}
\end{table*}
The Dermo-DOCTOR generates the best segmentation results for three out of the six cases while performs as second-best for the remaining three cases.  
Our Dermo-DOCTOR outperforms the DCL-PSI \citep{bi2019step} for ISIC-2016 by a margin of $1.0\,\%$ for mSp, whereas it has been lost by $1.0\,\%$ concerning mRc of DCL-PSI, while mIoU is constant for both methods. 
The current method (HRFB) of \citet{xie2020skin} has $5.0\,\%$ and $1.0\,\%$ less mRc and mSp, respectively, for ISIC-2016, while the mIoU is the same as our proposed method. 
It reveals that the proposed Dermo-DOCTOR has $5.0\,\%$ and $1.0\,\%$ less type-II and type-I errors, respectively, in the segmented lesion masks than HRFB.
The Dermo-DOCTOR also defeats the other two methods, the FCN ensemble \citep{yuan2017automatic} and Fusion Structure \citep{tang2019multi}, for all metrics and mSp for the ISIC-2016 dataset. 
Moreover, the methods in Table~\ref{tab:compareSEG} for the ISIC-2017 dataset have been defeated by the proposed Dermo-DOCTOR for mIoU, where it beats the nearby HRFB and iFCN \citep{ozturk2020skin} by a margin of $2.0\,\%$. 
Regarding the mRc and mSp, the Dermo-DOCTOR has been lost by $1.0\,\%$ by the methods in \citep{hasan2020dsnet,ozturk2020skin,xie2020skin} (see in Table~\ref{tab:compareSEG}). Still, the Dermo-DOCTOR performs better as it has $2.0\,\%$ more mIoU than the nearby methods \citep{ozturk2020skin,xie2020skin} for the ISIC-2107 dataset.

However, the above-discussions quantitatively and qualitatively demonstrate that the proposed Dermo-DOCTOR yields the most reliable skin lesion's ROIs, which proves its supremacy for lesion segmentation. Consequently, we will further utilize the proposed masks for lesion ROI extraction to perform the lesion recognition in the next subsection.

\subsection{Results for Recognition}
\label{classification}
This subsection exhibits the quantitative and qualitative results for lesion recognition, applying the proposed Dermo-DOCTOR and two other implemented well-known networks: the ResNet-$50$ and the Xception. In the end, we compare our results with several state-of-the-art results for those datasets. 
We utilize recall,  precision, and F1-score to quantify the recognition efficiency, where they respectively quantify the type-II error, the positive predictive values, and the harmonic mean of recall and precision for revealing the trade-off between them. 
Additionally, we also estimate the ROC curve and its corresponding AUC value to evaluate any randomly elected query image's prediction probability.

Table~\ref{tab:allresults} gives the lesion recognition results, showing the outcomes for three networks and two datasets. 
\begin{table*}[!ht]
\caption{The recognition results from numerous comprehensive experiments on the ISIC-$2016$ and ISIC-$2017$ test datasets on three separate networks, highlighting the weighted average metrics' highest values.}
\footnotesize
\begin{tabular}{ccc|cc|cc|cc}
\hline
\rowcolor[HTML]{C0C0C0} 
\cellcolor[HTML]{C0C0C0}                               & \multicolumn{2}{c|}{\cellcolor[HTML]{C0C0C0}}                                     & \multicolumn{2}{c}{\cellcolor[HTML]{C0C0C0}ResNet-$50$}      & \multicolumn{2}{|c|}{\cellcolor[HTML]{C0C0C0}Xception}      & \multicolumn{2}{c}{\cellcolor[HTML]{C0C0C0}Dermo-DOCTOR} \\ \cline{4-9} 
\rowcolor[HTML]{C0C0C0} 
\cellcolor[HTML]{C0C0C0}                               & \multicolumn{2}{c|}{\cellcolor[HTML]{C0C0C0}}                                     & \multicolumn{2}{c}{\cellcolor[HTML]{C0C0C0}Preprocessing} & \multicolumn{2}{|c|}{\cellcolor[HTML]{C0C0C0}Preprocessing} & \multicolumn{2}{c}{\cellcolor[HTML]{C0C0C0}Preprocessing}      \\ \cline{4-9} 
\rowcolor[HTML]{C0C0C0} 
\multirow{-3}{*}{\cellcolor[HTML]{C0C0C0}Test Dataset} & \multicolumn{2}{c|}{\multirow{-3}{*}{\cellcolor[HTML]{C0C0C0}Class-wise Metrics}} & $P_1$                                     & $P_2$                 & $P_1$                                     & $P_2$                & $P_1$                              & $P_2$                            \\ \hline
                                                       & \multicolumn{1}{c}{}                                        & Nev                & \multicolumn{1}{c}{}   $0.94$               &   0.87                & \multicolumn{1}{c}{}   $0.97$               &   $0.91$                & \multicolumn{1}{c}{} $0.96$        &  $0.90$                             \\ \cline{3-9} 
                                                       & \multicolumn{1}{c}{}                                        & Mel                & \multicolumn{1}{c}{}            $0.45$     &     $0.93$              & \multicolumn{1}{c}{}   $0.25$               & $0.92$                  & \multicolumn{1}{c}{}   0.37       &  $0.95$                             \\ \cline{3-9} 
                                                       & \multicolumn{1}{c}{\multirow{-3}{*}{Recall}}                & W. Avg.            & \multicolumn{1}{c}{}   $0.84$               &   $0.88$                & \multicolumn{1}{c}{}   $0.83$               &   $0.91$                & \multicolumn{1}{c}{} $0.84$         &    $\mathbf{0.91}$                           \\ \cline{2-9} 
                                                       & \multicolumn{1}{c}{}                                        & Nev                & \multicolumn{1}{c}{}   $0.87$              &    $0.98$                & \multicolumn{1}{c}{}   $0.84$               & $0.98$                 & \multicolumn{1}{c}{}   $0.86$       &    $0.99$                           \\ \cline{3-9} 
                                                       & \multicolumn{1}{c}{}                                        & Mel                & \multicolumn{1}{c}{}   $0.64$              &     $0.63$              & \multicolumn{1}{c}{}    $0.70$              &   $0.71$                & \multicolumn{1}{c}{}  $0.68$        &    $0.70$                           \\ \cline{3-9} 
                                                       & \multicolumn{1}{c}{\multirow{-3}{*}{Precision}}             & W. Avg.            & \multicolumn{1}{c}{}   $0.83$              &    $0.91$                & \multicolumn{1}{c}{}   $0.81$               &   $0.93$                & \multicolumn{1}{c}{}  $0.83$        &     $\mathbf{0.93}$                           \\ \cline{2-9} 
                                                       &                                                              & Nev                & \multicolumn{1}{c}{}   $0.90$              &   $0.92$                & \multicolumn{1}{c}{}     $0.90$             &   $0.94$                 & \multicolumn{1}{c}{}  0.91        &   $0.94$                            \\ \cline{3-9} 
                                                       &                                                              & Mel                & \multicolumn{1}{c}{} $0.53$                &     $0.75$              & \multicolumn{1}{c}{}      $0.37$            &   $0.80$                & \multicolumn{1}{c}{} $0.48$         &   $0.80$                            \\ \cline{3-9} 
\multirow{-9}{*}{ISIC-2016}                            & \multirow{-3}{*}{F1-score}                                   & W. Avg.            & \multicolumn{1}{c}{}       $0.83$          &         $0.89$          & \multicolumn{1}{c}{}       $0.80$           &   $0.91$                  & \multicolumn{1}{c}{}     $0.82$     &  $\mathbf{0.91}$       \\ \hline  \hline

                                                       & \multicolumn{1}{c}{}                                        & Nev                & \multicolumn{1}{c}{}             $0.80$    &   $0.80$               & \multicolumn{1}{c}{}           $0.86$       &  $0.80$                  & \multicolumn{1}{c}{}     $0.87$      &  $0.82$                              \\ \cline{3-9} 
                                                       & \multicolumn{1}{c}{}                                        & SK                 & \multicolumn{1}{c}{}             $0.59$    &   $0.80$               & \multicolumn{1}{c}{}            $0.66$      &    $0.83$                & \multicolumn{1}{c}{}    $0.66$       &    $0.82$                          \\ \cline{3-9} 
                                                       & \multicolumn{1}{c}{}                                        & Mel                & \multicolumn{1}{c}{}             $0.45$    &   $0.59$                & \multicolumn{1}{c}{}        $0.42$           &    $0.53$               & \multicolumn{1}{c}{}     $0.50$     &  $0.62$                             \\ \cline{3-9} 
                                                       & \multicolumn{1}{c}{\multirow{-4}{*}{Recall}}                & W. Avg.            & \multicolumn{1}{c}{}               $0.70$  &   $0.76$                & \multicolumn{1}{c}{}          $0.74$        &  $0.75$                 & \multicolumn{1}{c}{}    $0.77$      &    $\mathbf{0.78}$                           \\ \cline{2-9} 
                                                       & \multicolumn{1}{c}{}                                        & Nev                & \multicolumn{1}{c}{}             $0.80$    &  $0.88$                 & \multicolumn{1}{c}{}    $0.83$              &   $0.88$                & \multicolumn{1}{c}{}   $0.84$       &           $0.89$                   \\ \cline{3-9} 
                                                       & \multicolumn{1}{c}{}                                        & SK                 & \multicolumn{1}{c}{}             $0.64$    &   $0.59$               & \multicolumn{1}{c}{}  $0.61$                 &    $0.52$              & \multicolumn{1}{c}{}    $0.65$       &  $0.65$                           \\ \cline{3-9} 
                                                       & \multicolumn{1}{c}{}                                        & Mel                & \multicolumn{1}{c}{}             $0.43$    &  $0.56$                & \multicolumn{1}{c}{}  $0.52$                & $0.62$                  & \multicolumn{1}{c}{}    $0.58$       &  $0.57$                            \\ \cline{3-9} 
                                                       & \multicolumn{1}{c}{\multirow{-4}{*}{Precision}}             & W. Avg.            & \multicolumn{1}{c}{}            $0.70$     &   $0.78$              & \multicolumn{1}{c}{}  $0.74$                &   $0.77$                 & \multicolumn{1}{c}{}  $0.76$     &   $\mathbf{0.79}$                           \\ \cline{2-9} 
                                                       & \multicolumn{1}{c}{}                                        & Nev                & \multicolumn{1}{c}{}             $0.80$    &  0.84             & \multicolumn{1}{c}{}  0.85                &  $0.84$                  & \multicolumn{1}{c}{}     $0.85$         &  $0.85$                           \\ \cline{3-9} 
                                                       & \multicolumn{1}{c}{}                                        & SK                 & \multicolumn{1}{c}{}            $0.61$     &  $0.68$                & \multicolumn{1}{c}{}  $0.63$                &     $0.64$               & \multicolumn{1}{c}{} $0.65$            &  $0.73$                             \\ \cline{3-9} 
                                                       & \multicolumn{1}{c}{}                                        & Mel                & \multicolumn{1}{c}{}            $0.44$     &   $0.57$               & \multicolumn{1}{c}{}  $0.46$                  &   $0.57$                 & \multicolumn{1}{c}{}     $0.53$       & $0.59$                                \\ \cline{3-9} 
\multirow{-12}{*}{ISIC-2017}                           & \multicolumn{1}{c}{\multirow{-4}{*}{F1-score}}              & W. Avg.            & \multicolumn{1}{c}{}             $0.70$    &     $0.76$             & \multicolumn{1}{c}{}  $0.74$                 &       $0.76$             & \multicolumn{1}{c}{}    $0.76$       &      $\mathbf{0.78}$                         \\ \hline

\multicolumn{9}{l}{\scriptsize{$P_1$: Segmentation; $P_2$: Segmentation+Rebalancing+Augmentation; W. Avg.: Weighted Average}}
\end{tabular}
\label{tab:allresults}
\end{table*}
The weighted average of recalls for ResNet-$50$, Xception, and Dermo-DOCTOR have been respectively improved by the margins of $4.0\,\%$, $8.0\,\%$, and $7.0\,\%$ for ISIC-2016, when we employ the preprocessing $P_2$ instead of baseline $P_1$ (see in Table~\ref{tab:allresults}).
The highest possible recall ($0.91$) for ISIC-2016 is received from the proposed Dermo-DOCTOR classifier, applying the proposed preprocessing $P_2$ on the segmented masks from the proposed Dermo-DOCTOR segmentor. 
The rebalancing and augmentation, along with the segmentation, in preprocessing $P_2$ reduces the FN-rates ($63.0\,\%$ to $5.0\,\%$) of Mel class. 
It also decreases the FP-rates ($32.0\,\%$ to $30.0\,\%$) for the same class while applying our proposed Dermo-DOCTOR. 
Such reductions, in FN- and FP-rates, are praiseworthy of our proposed preprocessing $P_2$ and Dermo-DOCTOR comparing the baseline classifiers (ResNet-$50$ and Xception) and preprocessing $P_1$.
Likewise, the weighted average of precision and F1-score on ISIC-2016 are also respectively improved by $10.0\,\%$ and $9.0\,\%$ for the preprocessing $P_2$. 
Remarkably, the harmonic mean of recall and precision for the Mel class of ISIC-2016 has significantly strengthened by a margin of $32.0\,\%$ with preprocessing ($P_2$) and Dermo-DOCTOR. 
However, comparing all the experiments on the ISIC-2016, the proposed $P_2$ and Dermo-DOCTOR are the best preprocessing and classifier having a type-II error of $9.0\,\%$ and positive predictive value of $93.0\,\%$. 
Table~\ref{tab:allresults} also notes that the weighted average of recall, precision, and F1-score have the respective maximum values of $0.78$, $0.79$, and $0.78$, for the preprocessing $P_2$ and proposed Dermo-DOCTOR on the ISIC-2017. 
The recall of SK and Mel classes in the ISIC-2017 dataset, applying the proposed Dermo-DOCTOR and preprocessing $P_2$, have respectively updated by the margins of $16.0\,\%$ and $12.0\,\%$. In comparison, it decreases by $1.0\,\%$ for the Nev class, which is acceptable in the medical diagnostic system (as the positive class is significantly improved). 
A margin of $5.0\,\%$ has raised the Nev class's positive predictive value, whereas it is closer or equal for the other two classes for the proposed Dermo-DOCTOR with $P_2$.
However, comparing all the experiments on the ISIC-2017, the proposed $P_2$ and Dermo-DOCTOR are the best preprocessing and classifier, with a type-II error of $22.0\,\%$ and positive predictive value of $79.0\,\%$. 

The experimental lesion categorization results on ISIC-2016 and ISIC-2017 datasets demonstrate that the two-classes' (ISIC-2016) recognition performance is better than the three-classes (ISIC-2017). 
The addition of SK class in ISIC-2017 reduces the weighted average recall, precision, and F1-score by the margins of $13.0\,\%$, $14.0\,\%$, and $13.0\,\%$, respectively. 
The higher similarity of SK with Nev and Mel classes is the possibility of such reduced ISIC-2017 test results. Noticeably, more classes tend to bring complications to the classifiers, especially when the training has fewer examples and inter-class similarities. The joining of different heterogeneous samples to each class can enhance the categorization results in ISIC-2017.

\begin{figure*}[!ht]
  \centering
  \subfloat[]{\includegraphics[width=8cm, height= 5.7cm]{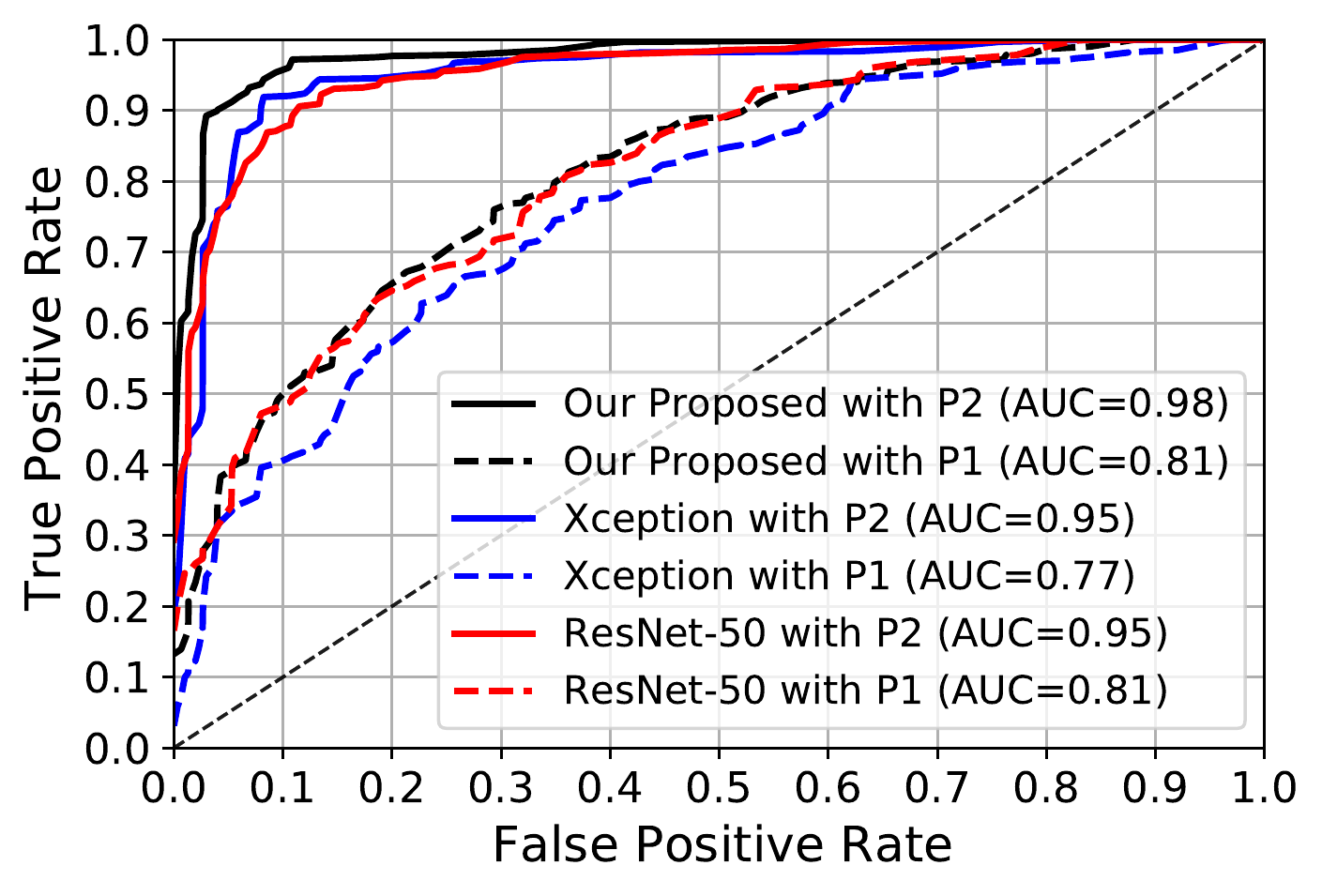}} \hspace{0.25cm}
  \subfloat[]{\includegraphics[width=8cm, height= 5.7cm]{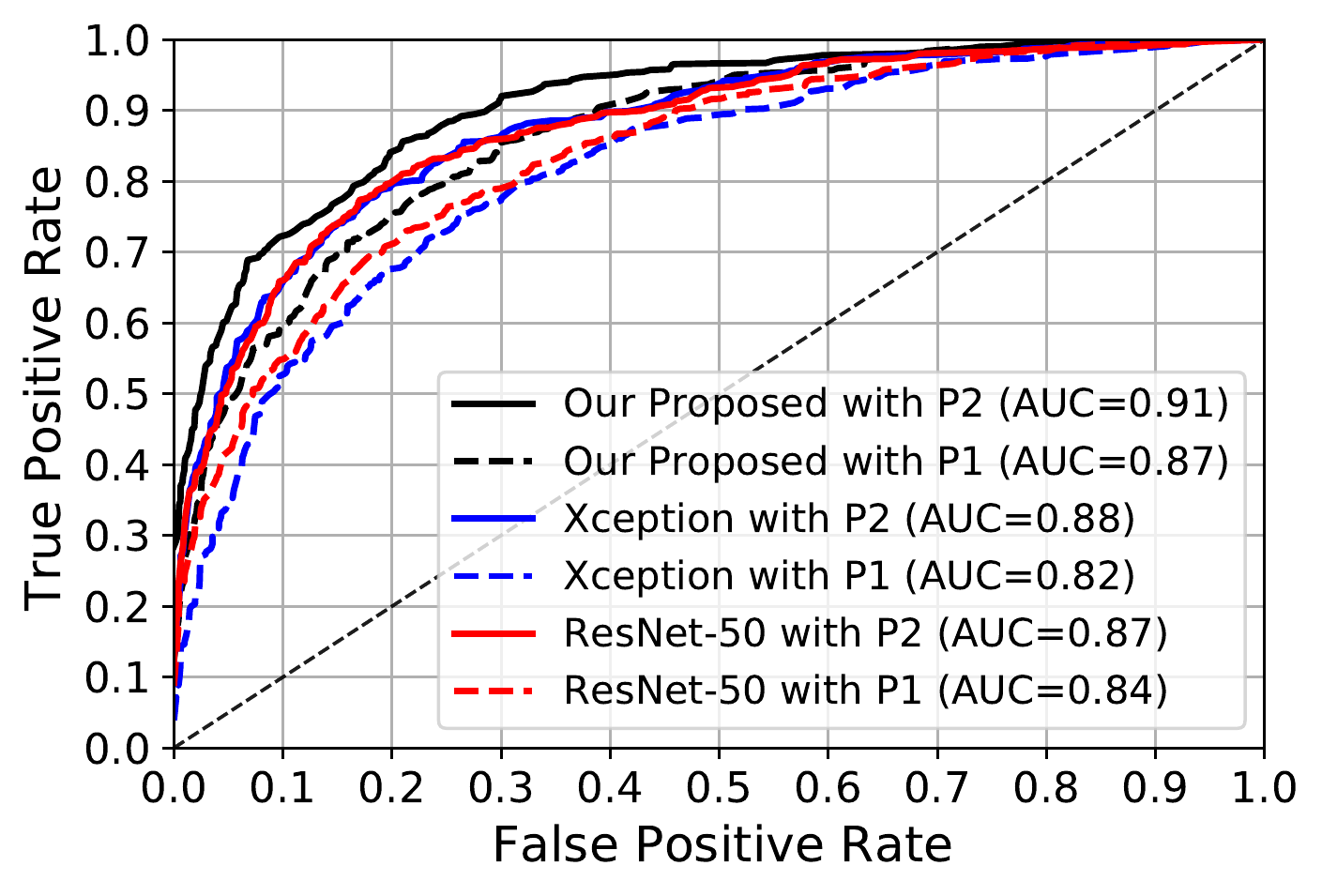}} 
  \caption{The ROC curves ((a) for ISIC-2016 and (b) for ISIC-2017) of skin lesion recognition, where we have plotted the ROC for the proposed Dermo-DOCTOR and implemented ResNet-$50$ and Xception with two different preprocessing ($P_1$ and $P_2$).}
  \label{fig:ROC}
\end{figure*}

The further analysis of different classifiers and preprocessing have presented in the ROC curves in Fig.~\ref{fig:ROC}, showing the highest possible AUC as $0.98$ and $0.91$ respectively for ISIC-2016 and ISIC-2017.
The rebalancing and augmentation employment with segmentation ($P_2$) heightens AUC for both ISIC-$2016$ and ISIC-$2017$ datasets of all classifiers.
The Dermo-DOCTOR with $P_2$ has beaten the baseline Xception and ResNet-$50$ respectively by $3.0\,\%$ and $3.0\,\%$ in terms of AUC for the ISIC-2016 dataset. Whereas it surpasses them by the margins of $3.0\,\%$ and $4.0\,\%$ for AUC for the ISIC-2017 dataset. 
However, the earlier discussions for the lesion recognition on ISIC-$2016$ and ISIC-$2017$ test datasets expose the supremacy of the proposed Dermo-DOCTOR and preprocessing $P_2$.  

\begin{figure*}[!ht]  
\centering  
\includegraphics[width=16cm, height= 14cm]{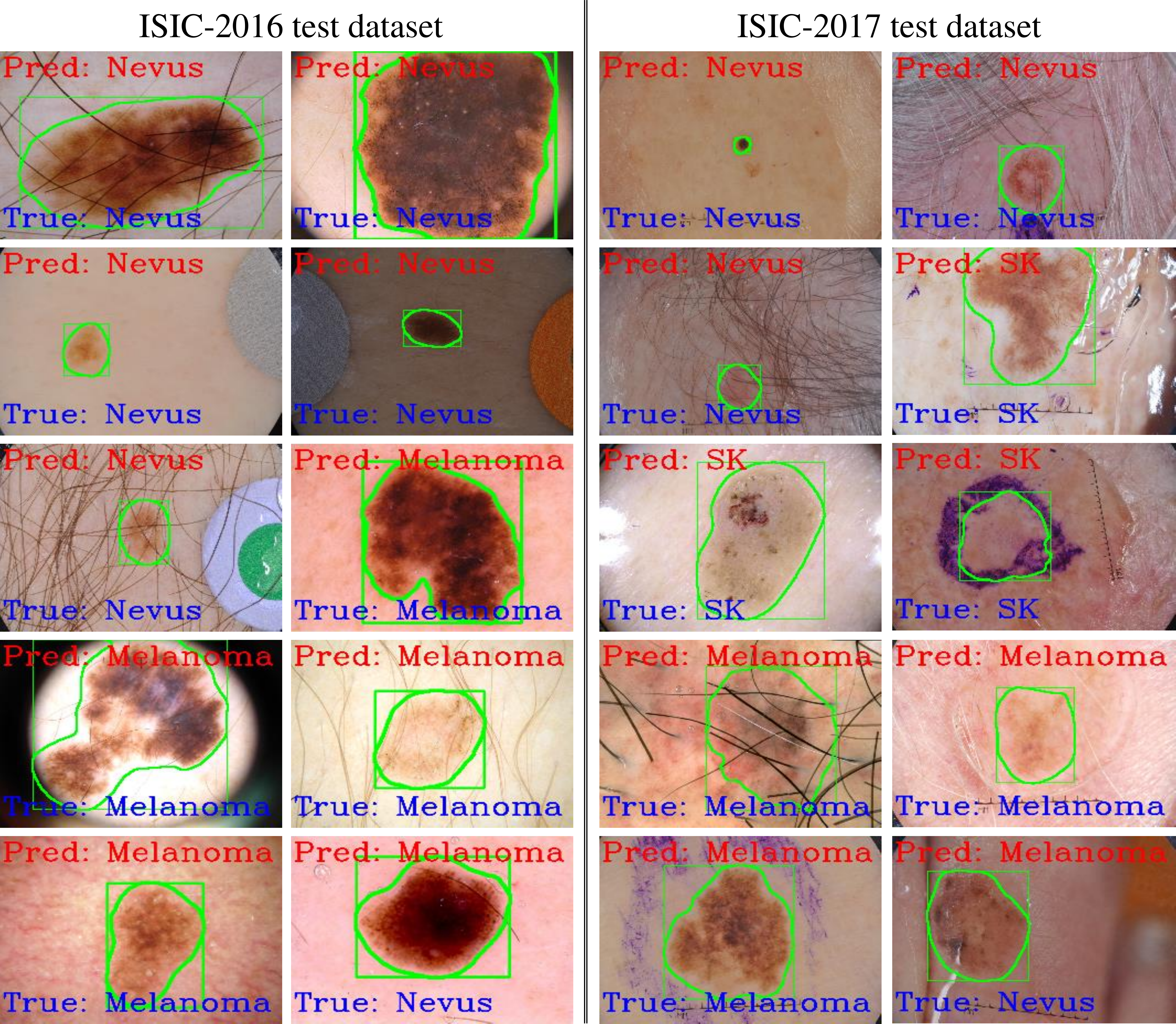}
\caption{Example of several qualitative classification results of the challenging images of the ISIC-2016 and ISIC-2017 test datasets using the Dermo-DOCTOR, where the recognition has accomplished using the segmented ROIs (green color) from the Dermo-DOCTOR.}
\label{fig:qualiResults}
\end{figure*}

The detailed class-wise performances of the lesion recognition by the proposed Dermo-DOCTOR and preprocessing $P_2$ are exhibited in Table~\ref{tab:confusion}.  
\begin{table*}[!ht]
\centering
\caption{The confusion matrix for the ISIC-$2016$ with 379 samples (left) and ISIC-$2016$ with 600 samples (right) test datasets, employing the proposed Dermo-DOCTOR and preprocessing $P_2$}
\begin{tabular}{ll}
\begin{tabular}{cccc}
\hline
                            &     & \multicolumn{2}{c}{Actual}                          \\ \cline{3-4} 
                            &     & Nev                      & Mel                      \\ \hline
                            & Nev & \cellcolor[HTML]{C0C0C0} \begin{tabular}[c]{@{}l@{}}273 \\ $89.80\,\%$ \end{tabular} &  \begin{tabular}[c]{@{}l@{}}4 \\ $5.33\,\%$ \end{tabular}                         \\ \cline{2-4} 
\multirow{-2}{*}{\rotatebox[origin=c]{90}{Predicted}} & Mel &              \begin{tabular}[c]{@{}l@{}}31 \\ $10.20\,\%$ \end{tabular}            & \cellcolor[HTML]{C0C0C0} \begin{tabular}[c]{@{}l@{}}71 \\ $94.67\,\%$ \end{tabular} \\ \hline
\end{tabular}
&
\begin{tabular}{ccccc}
\hline
                            &     & \multicolumn{3}{c}{Actual}                                                     \\ \cline{3-5} 
                            &     & Nev                      & SK                       & Mel                      \\ \hline
                            & Nev & \cellcolor[HTML]{C0C0C0} \begin{tabular}[c]{@{}l@{}}321 \\ $81.68\,\%$ \end{tabular} &    \begin{tabular}[c]{@{}l@{}}10 \\ $11.11\,\%$ \end{tabular}                       &   \begin{tabular}[c]{@{}l@{}}28 \\ $23.93\,\%$ \end{tabular}                        \\ \cline{2-5} 
                            & SK  &      \begin{tabular}[c]{@{}l@{}}23 \\ $5.85\,\%$ \end{tabular}                     & \cellcolor[HTML]{C0C0C0} \begin{tabular}[c]{@{}l@{}}74 \\ $82.22\,\%$ \end{tabular} &   \begin{tabular}[c]{@{}l@{}}17 \\ $14.53\,\%$ \end{tabular}                        \\ \cline{2-5} 
\multirow{-3}{*}{\rotatebox[origin=c]{90}{Predicted}} & Mel &     \begin{tabular}[c]{@{}l@{}}49 \\ $12.47\,\%$ \end{tabular}                      &        \begin{tabular}[c]{@{}l@{}}6 \\ $6.67\,\%$ \end{tabular}                   & \cellcolor[HTML]{C0C0C0} \begin{tabular}[c]{@{}l@{}}72 \\ $61.54\,\%$ \end{tabular} \\ \hline
\end{tabular}
\end{tabular}
\label{tab:confusion}
\end{table*}
The  ISIC-2016's  confusion matrix in Table~\ref{tab:confusion} (left) shows that among 304-Nev samples correctly recognized samples are $273\,(89.80\,\%)$, whereas barely $31\,(10.20\,\%)$-Nev samples are recognized as Mel (as FP). 
It also reveals that among 75-Mel samples, rightly recognized samples are $71\,(94.67\,\%)$, whereas only $4\,(5.33\,\%)$-Mel samples are improperly recognized as Nev (as FN). 
Again, the ISIC-2017's confusion matrix (see in Table~\ref{tab:confusion} (right)) demonstrates that $81.68\,\%$-Nev samples are correctly recognized as Nev class, while $18.32\,\%$-Nev  samples are wrongly recognized to other classes as FP ($5.85\,\%$ as SK and $12.47\,\%$ as Mel). 
Similarly, $17.78\,\%$ and $38.46\,\%$ samples of the SK- and Mel-classes belong to FP and FN, respectively. 
Although the $38.46\,\%$ of the positive samples (Mel) are improperly recognized, it is still better than the baseline $58.0\,\%$ errors in the baseline (Xception with preprocessing $P_1$).     
Fig.~\ref{fig:qualiResults} bestows qualitative results from the proposed Dermo-DOCTOR classifier and preprocessing $P_2$  for the lesion recognition into different, either two classes or three classes. 
For concurrent detection and recognition, we utilize the segmented masks and categorized class for contouring the lesions (green color) and label annotation on the image to help the dermatologists for further assessment (see in Fig.~\ref{fig:qualiResults}). More concurrent results for all the test images are available on YouTube (ISIC-$2016$\footnote{ISIC-$2016$ (Detection \& Recognition): \url{https://bit.ly/Dermo-DOCTOR_ISIC_16}} and ISIC-$2017$\footnote{ISIC-$2017$ (Detection \& Recognition): \url{https://bit.ly/Dermo-DOCTOR_ISIC_17}}). 
However, the results in Fig.~\ref{fig:qualiResults} illustrate a few challenging images, where we show some wrongly recognized images. 
Those qualitative results depict that the detection and recognition are precise even the query test images contain different artifacts (see in Fig.~\ref{fig:artifacts}). 
Although the Dermo-DOCTOR incorrectly predicts some images, they visually seem like a predicted class.

Table~\ref{tab:previousWork2016} describes the comparison of the results of our Dermo-DOCTOR and other methods, which were trained and tested on the same ISIC datasets.
The proposed Dermo-DOCTOR produces the best recognition for two out of the six cases while performing second-best with the winning methods on the other four cases (see in Table~\ref{tab:previousWork2016}).
\begin{table*}[!ht]
\footnotesize
\caption{The state-of-the-art comparison with proposed Dermo-DOCTOR, which had trained, validated, and tested on the ISIC-$2016$ and ISIC-$2017$ datasets.}
\begin{tabular}{l|ccc|ccc}
\hline
\rowcolor[HTML]{C0C0C0} 
\cellcolor[HTML]{C0C0C0}                          & \multicolumn{3}{c|}{\cellcolor[HTML]{C0C0C0}ISIC-2016 test dataset} & \multicolumn{3}{c}{\cellcolor[HTML]{C0C0C0}ISIC-2017 test dataset}  \\ \cline{2-7} 
\rowcolor[HTML]{C0C0C0} 
\multirow{-2}{*}{\cellcolor[HTML]{C0C0C0}Classification Methods} & Recall           & Precision           & AUC          & Recall           & Precision           & AUC                    \\ \hline

ResNet-$50$ \citep{brinker2019enhanced}                                           &  $0.56$                &    $0.71$                 &     $0.85$         &       -           &           -          &  -                    \\

GR \citep{serte2019gabor} &        -          &         -            &      -        &          $0.15$        &         -            &    $0.91$          \\

ARLCNN \citep{zhang2019attention} &    -              &      -               &      -        &    $0.77$              &          -           &    $\mathbf{0.92}$           \\

IR \citep{al2020multiple}                                                &   $0.82$               &   -                  &  $0.77$            &    $0.76$              &         -            &    -                  \\

FPRPN \citep{song2020end}                                             &  \boldmath{$0.99$}                &    $\underline{0.82}$                 &     $0.81$         &     \boldmath{$0.98$}             &          $\mathbf{0.82}$           &  $0.79$                    \\ 

MFA \citep{yu2020convolutional}                                         &  $0.60$                &    $0.69$                 &     $\underline{0.86}$         &         -         &         $0.72$            &      $0.90$           \\


\textbf{Proposed Dermo-DOCTOR (2020)} &    $\underline{0.91}$             &         \boldmath{$0.93$}           &     \boldmath{$0.98$}         &  $\underline{0.78}$                & $\underline{0.79}$                   &     $\underline{0.91}$             \\ \hline 

\multicolumn{7}{l}{\scriptsize{\textbf{GR:} Gabor Wavelet-based CNN
\citep{serte2019gabor}}} \\

\multicolumn{7}{l}{\scriptsize{\textbf{ARLCNN:} Attention Residual Learning CNN (ResNet-$14$ \& ResNet-$50$) \citep{zhang2020skin}}} \\

\multicolumn{7}{l}{\scriptsize{\textbf{IR:} Inception-ResNet-V$2$ (ISIC-$2016$), ResNet-$50$ (ISIC-$2017$,ISIC-$2018$) \citep{al2020multiple}}} \\ 

\multicolumn{7}{l}{\scriptsize{\textbf{FPRPN:} Feature Pyramid Network (FPN) and Region Proposal Network (RPN) \citep{song2020end}}} \\


\multicolumn{7}{l}{\scriptsize{\textbf{MFA:} Multi-network based feature aggregation \citep{yu2020convolutional}}} \\

\end{tabular}
\label{tab:previousWork2016}
\end{table*}

\textbf{Comparison of ISIC-2016.} 
The proposed network produces the best results for the AUC by beating the state-of-the-art of \citet{yu2020convolutional} with a $12.0\,\%$ border.
Concerning the type-II errors, Dermo-DOCTOR is behind the state-of-the-art (FPRPN) of \citet{song2020end} by a border of $8.0\,\%$, but the Dermo-DOCTOR outperforms it by a $11.0\,\%$ border for the positive predictive value. 
However, Dermo-DOCTOR beats the FPRPN by a $1.5\,\%$ perimeter in terms of balanced accuracy. Whereas Dermo-DOCTOR also exceeds the FPRPN by a $17.0\,\%$ margin for AUC.  

\textbf{Comparison of ISIC-2017.}
The Dermo-DOCTOR serves as the second-best results concerning all the metrics, where it beats the state-of-the-art FPRPN \citep{song2020end} with a margin of $12.0\,\%$ for AUC, although FPRPN has defeated it for balanced accuracy. 
The proposed Dermo-DOCTOR has defeated ARLCNN \citep{zhang2019attention} by a $1.0\,\%$ border concerning the recall, but ARLCNN wins by a $1.0\,\%$ margin in AUC.  
However, the proposed Dermo-DOCTOR produces the second-highest results for recall by beating the third-best ARLCNN \citep{zhang2019attention} with a margin of $1.0\,\%$. 
For precision, it performs so by vanquishing third-best MFA \citep{yu2020convolutional} with a border of $7.0\,\%$, and for AUC by beating third-best MFA with a margin of $1.0\,\%$.

\subsection{Applications}
\begin{figure*}[!ht]
  \centering
  \subfloat{\includegraphics[width=16cm, height= 8cm]{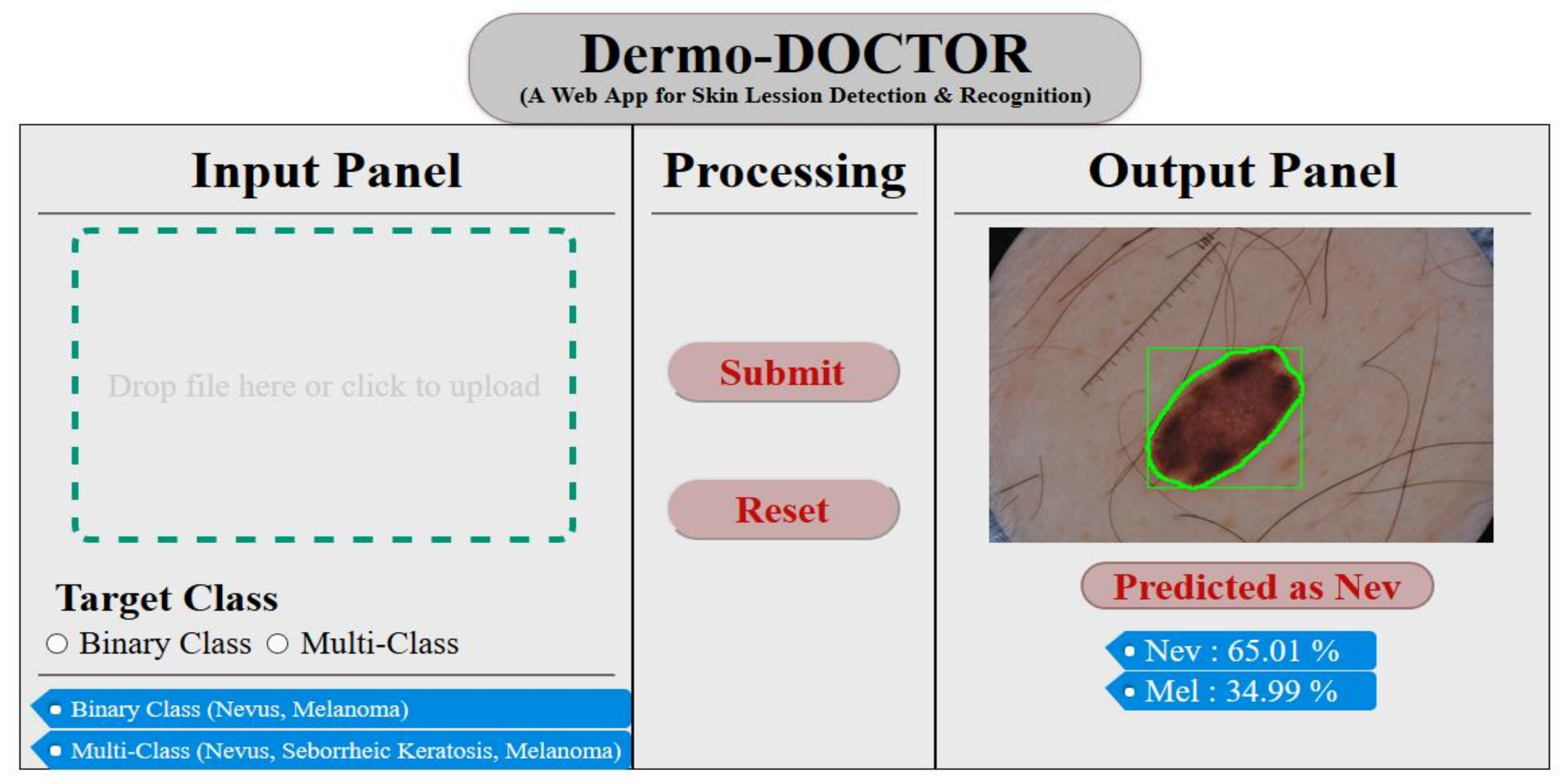}}
  \caption{Our designed web application to detect and recognize skin lesions simultaneously. Dermatologists can utilize this application by selecting or dragging the image as an input. The app will confer the lesion detection and recognition results for either binary class or multi-classes.}
  \label{fig:dermoApplication}
\end{figure*}

Fig.~\ref {fig:dermoApplication} illustrates the prototype of the developed web application deploying our Dermo-DOCTOR, which runs in a web browser at \say{http://127.0.0.1:5000/} by accessing the CNN environments of the local machine. 
The app takes a dermoscopic image (png, jpg, bmp, or jpeg) as an input, displaying the user's diagnosis result as soon as the back-end model analyzes a given image. 
The recognized class with its probability is displayed in the output panel, highlighting the lesion ROI by a green color bounding box. 
Therefore, it helps the dermatologists focus on that detected area for cross-checking the predicted class. 
A real-time utilization of the Dermo-DOCTOR has been uploaded to YouTube\footnote{Dermo-DOCTOR App: \url{https://bit.ly/Dermo-DOCTOR_App}}, which confers less time-latency of getting results. 
The reason for that is the app sends images to the host and receives the host's results through the internet and the time for prediction in the host (higher traffic in the host will increase the latency).
However, the dedicated machine with GPU can alleviate this time-latency limitation. We have tested the app on our local machine and could not make it public due to resource limitations.

\section{Conclusion}
\label{Conclusion}
Despite the present colossal challenges due to high visual and intra-class variability, inter-class similarity, and the presence of different artifacts, automated skin lesion detection and recognition are incredibly crucial.
However, this article proposed and developed an automated CNN-based lesion detection and recognition network, integrating ROI extraction by segmentation, image augmentations, and class rebalancing.
Our experimental results demonstrated that the proposed Dermo-DOCTOR could detect and recognize the lesion more accurately as we concatenated features from two different encoders. 
Such a concatenation provides more prominent and discriminating feature maps of the skin lesions comparing a single encoder.
The segmented lesions rather than the whole images can provide more salient and representative features from the CNNs, leading to improved lesion recognition. 
Moreover, the rebalanced class distribution attained better performance of the recognition as compared to the imbalanced distribution.
Additionally, the augmentation led the CNN-based classifier to be more generic as CNNs can learn from diverse training samples.
Thus, it achieved state-of-the-art performance to detect and recognize the lesions from two different test datasets, such as ISIC-$2016$ and ISIC-$2017$.
We will further explore and investigate the effects of improved segmentation and weighting of the underrepresented classes in the future.
The deployment of our framework to a web application precisely detected and recognized the lesions concurrently. 
The developed web application will be improved, making it more user-friendly for dermatologists and deploying it to the google cloud platform for clinical applications.

\section*{Author Contributions}
\textbf{M. K. Hasan:} Conceptualization, Methodology, Software, Formal analysis, Investigation, Writing- Review \& Editing, Supervision;
\textbf{S. Roy:} Validation, Data Curation, Writing- Original Draft;
\textbf{C. Mondal:} Validation, Data Curation, Writing- Original Draft;
\textbf{M. A. Alam:} Conceptualization, Data Curation;
\textbf{M. T. E. Elahi:} Validation, Writing- Original Draft;
\textbf{A. Dutta:} Validation, Writing- Original Draft;
\textbf{S. M. T. U. Raju:} Software;
\textbf{M. T. Jawad:} Validation, Writing- Review \& Editing;
\textbf{M. Ahmad:} Supervision.

\section*{Acknowledgements}
None. No funding to declare. 

\section*{Conflict of Interest}
All authors have no conflict of interest to publish this research. 

\bibliographystyle{model2-names}
\bibliography{sample}

\begin{thebibliography}{92}
\expandafter\ifx\csname natexlab\endcsname\relax\def\natexlab#1{#1}\fi
\expandafter\ifx\csname url\endcsname\relax
  \def\url#1{\texttt{#1}}\fi
\expandafter\ifx\csname urlprefix\endcsname\relax\def\urlprefix{URL }\fi
\providecommand{\eprint}[2][]{\url{#2}}
\providecommand{\bibinfo}[2]{#2}
\ifx\xfnm\relax \def\xfnm[#1]{\unskip,\space#1}\fi
\bibitem[{Acharya et~al.(2012)Acharya, Molinari, Sree, Chattopadhyay, Ng and
  Suri}]{acharya2012automated}
\bibinfo{author}{Acharya, U.R.}, \bibinfo{author}{Molinari, F.},
  \bibinfo{author}{Sree, S.V.}, \bibinfo{author}{Chattopadhyay, S.},
  \bibinfo{author}{Ng, K.H.}, \bibinfo{author}{Suri, J.S.},
  \bibinfo{year}{2012}.
\newblock \bibinfo{title}{Automated diagnosis of epileptic eeg using
  entropies}.
\newblock \bibinfo{journal}{Biomedical Signal Processing and Control}
  \bibinfo{volume}{7}, \bibinfo{pages}{401--408}.
\bibitem[{Al-Masni et~al.(2018)Al-Masni, Al-antari, Choi, Han and
  Kim}]{al2018skin}
\bibinfo{author}{Al-Masni, M.A.}, \bibinfo{author}{Al-antari, M.A.},
  \bibinfo{author}{Choi, M.T.}, \bibinfo{author}{Han, S.M.},
  \bibinfo{author}{Kim, T.S.}, \bibinfo{year}{2018}.
\newblock \bibinfo{title}{Skin lesion segmentation in dermoscopy images via
  deep full resolution convolutional networks}.
\newblock \bibinfo{journal}{Computer methods and programs in biomedicine}
  \bibinfo{volume}{162}, \bibinfo{pages}{221--231}.
\bibitem[{Al-Masni et~al.(2020)Al-Masni, Kim and Kim}]{al2020multiple}
\bibinfo{author}{Al-Masni, M.A.}, \bibinfo{author}{Kim, D.H.},
  \bibinfo{author}{Kim, T.S.}, \bibinfo{year}{2020}.
\newblock \bibinfo{title}{Multiple skin lesions diagnostics via integrated deep
  convolutional networks for segmentation and classification}.
\newblock \bibinfo{journal}{Computer Methods and Programs in Biomedicine}
  \bibinfo{volume}{190}, \bibinfo{pages}{105351}.
\bibitem[{Al~Nazi and Abir(2020)}]{al2020automatic}
\bibinfo{author}{Al~Nazi, Z.}, \bibinfo{author}{Abir, T.A.},
  \bibinfo{year}{2020}.
\newblock \bibinfo{title}{Automatic skin lesion segmentation and melanoma
  detection: Transfer learning approach with u-net and dcnn-svm}, in:
  \bibinfo{booktitle}{Proceedings of International Joint Conference on
  Computational Intelligence}, \bibinfo{organization}{Springer}. pp.
  \bibinfo{pages}{371--381}.
\bibitem[{Amin et~al.(2020)Amin, Sharif, Gul, Anjum, Nisar, Azam and
  Bukhari}]{amin2020integrated}
\bibinfo{author}{Amin, J.}, \bibinfo{author}{Sharif, A.}, \bibinfo{author}{Gul,
  N.}, \bibinfo{author}{Anjum, M.A.}, \bibinfo{author}{Nisar, M.W.},
  \bibinfo{author}{Azam, F.}, \bibinfo{author}{Bukhari, S.A.C.},
  \bibinfo{year}{2020}.
\newblock \bibinfo{title}{Integrated design of deep features fusion for
  localization and classification of skin cancer}.
\newblock \bibinfo{journal}{Pattern Recognition Letters} \bibinfo{volume}{131},
  \bibinfo{pages}{63--70}.
\bibitem[{Badrinarayanan et~al.(2017)Badrinarayanan, Kendall and
  Cipolla}]{badrinarayanan2017segnet}
\bibinfo{author}{Badrinarayanan, V.}, \bibinfo{author}{Kendall, A.},
  \bibinfo{author}{Cipolla, R.}, \bibinfo{year}{2017}.
\newblock \bibinfo{title}{{SegNet: A} deep convolutional encoder-decoder
  architecture for image segmentation}.
\newblock \bibinfo{journal}{IEEE transactions on pattern analysis and machine
  intelligence} \bibinfo{volume}{39}, \bibinfo{pages}{2481--2495}.
\bibitem[{Bi et~al.(2019)Bi, Kim, Ahn, Kumar, Feng and Fulham}]{bi2019step}
\bibinfo{author}{Bi, L.}, \bibinfo{author}{Kim, J.}, \bibinfo{author}{Ahn, E.},
  \bibinfo{author}{Kumar, A.}, \bibinfo{author}{Feng, D.},
  \bibinfo{author}{Fulham, M.}, \bibinfo{year}{2019}.
\newblock \bibinfo{title}{Step-wise integration of deep class-specific learning
  for dermoscopic image segmentation}.
\newblock \bibinfo{journal}{Pattern recognition} \bibinfo{volume}{85},
  \bibinfo{pages}{78--89}.
\bibitem[{Bi et~al.(2017)Bi, Kim, Ahn, Kumar, Fulham and
  Feng}]{bi2017dermoscopic}
\bibinfo{author}{Bi, L.}, \bibinfo{author}{Kim, J.}, \bibinfo{author}{Ahn, E.},
  \bibinfo{author}{Kumar, A.}, \bibinfo{author}{Fulham, M.},
  \bibinfo{author}{Feng, D.}, \bibinfo{year}{2017}.
\newblock \bibinfo{title}{Dermoscopic image segmentation via multistage fully
  convolutional networks}.
\newblock \bibinfo{journal}{IEEE Transactions on Biomedical Engineering}
  \bibinfo{volume}{64}, \bibinfo{pages}{2065--2074}.
\bibitem[{Bray et~al.(2018)Bray, Ferlay, Soerjomataram, Siegel, Torre and
  Jemal}]{bray2018global}
\bibinfo{author}{Bray, F.}, \bibinfo{author}{Ferlay, J.},
  \bibinfo{author}{Soerjomataram, I.}, \bibinfo{author}{Siegel, R.L.},
  \bibinfo{author}{Torre, L.A.}, \bibinfo{author}{Jemal, A.},
  \bibinfo{year}{2018}.
\newblock \bibinfo{title}{Global cancer statistics 2018: Globocan estimates of
  incidence and mortality worldwide for 36 cancers in 185 countries}.
\newblock \bibinfo{journal}{CA: a cancer journal for clinicians}
  \bibinfo{volume}{68}, \bibinfo{pages}{394--424}.
\bibitem[{Brinker et~al.(2019)Brinker, Hekler, Enk and von
  Kalle}]{brinker2019enhanced}
\bibinfo{author}{Brinker, T.J.}, \bibinfo{author}{Hekler, A.},
  \bibinfo{author}{Enk, A.H.}, \bibinfo{author}{von Kalle, C.},
  \bibinfo{year}{2019}.
\newblock \bibinfo{title}{Enhanced classifier training to improve precision of
  a convolutional neural network to identify images of skin lesions}.
\newblock \bibinfo{journal}{PloS one} \bibinfo{volume}{14}.
\bibitem[{Cheng et~al.(2008)Cheng, Swamisai, Umbaugh, Moss, Stoecker, Teegala
  and Srinivasan}]{cheng2008skin}
\bibinfo{author}{Cheng, Y.}, \bibinfo{author}{Swamisai, R.},
  \bibinfo{author}{Umbaugh, S.E.}, \bibinfo{author}{Moss, R.H.},
  \bibinfo{author}{Stoecker, W.V.}, \bibinfo{author}{Teegala, S.},
  \bibinfo{author}{Srinivasan, S.K.}, \bibinfo{year}{2008}.
\newblock \bibinfo{title}{Skin lesion classification using relative color
  features}.
\newblock \bibinfo{journal}{Skin Research and Technology} \bibinfo{volume}{14},
  \bibinfo{pages}{53--64}.
\bibitem[{Chollet(2017)}]{chollet2017xception}
\bibinfo{author}{Chollet, F.}, \bibinfo{year}{2017}.
\newblock \bibinfo{title}{Xception: Deep learning with depthwise separable
  convolutions}, in: \bibinfo{booktitle}{Proceedings of the IEEE conference on
  computer vision and pattern recognition}, pp. \bibinfo{pages}{1251--1258}.
\bibitem[{Codella et~al.(2018)Codella, Gutman, Celebi, Helba, Marchetti, Dusza,
  Kalloo, Liopyris, Mishra, Kittler et~al.}]{codella2018skin}
\bibinfo{author}{Codella, N.C.}, \bibinfo{author}{Gutman, D.},
  \bibinfo{author}{Celebi, M.E.}, \bibinfo{author}{Helba, B.},
  \bibinfo{author}{Marchetti, M.A.}, \bibinfo{author}{Dusza, S.W.},
  \bibinfo{author}{Kalloo, A.}, \bibinfo{author}{Liopyris, K.},
  \bibinfo{author}{Mishra, N.}, \bibinfo{author}{Kittler, H.}, et~al.,
  \bibinfo{year}{2018}.
\newblock \bibinfo{title}{Skin lesion analysis toward melanoma detection: A
  challenge at the 2017 international symposium on biomedical imaging (isbi),
  hosted by the international skin imaging collaboration (isic)}, in:
  \bibinfo{booktitle}{2018 IEEE 15th International Symposium on Biomedical
  Imaging (ISBI 2018)}, \bibinfo{organization}{IEEE}. pp.
  \bibinfo{pages}{168--172}.
\bibitem[{Deng et~al.(2009)Deng, Dong, Socher, Li, Li and
  Fei-Fei}]{deng2009imagenet}
\bibinfo{author}{Deng, J.}, \bibinfo{author}{Dong, W.},
  \bibinfo{author}{Socher, R.}, \bibinfo{author}{Li, L.J.},
  \bibinfo{author}{Li, K.}, \bibinfo{author}{Fei-Fei, L.},
  \bibinfo{year}{2009}.
\newblock \bibinfo{title}{Imagenet: A large-scale hierarchical image database},
  in: \bibinfo{booktitle}{2009 IEEE conference on computer vision and pattern
  recognition}, \bibinfo{organization}{Ieee}. pp. \bibinfo{pages}{248--255}.
\bibitem[{{Dennis Schmid}(2018)}]{Europen17online}
\bibinfo{author}{{Dennis Schmid}}, \bibinfo{year}{2018}.
\newblock \bibinfo{title}{\emph{Number of dermatologists in selected European
  countries in 2015}}.
\newblock \bibinfo{note}{\url{https://tinyurl.com/y59xoc7n} [Accessed: 4 Sept
  2020]}.
\bibitem[{{Department of Health (Commonwealth of
  Australia)}(2017)}]{Dermatology16Factsheet}
\bibinfo{author}{{Department of Health (Commonwealth of Australia)}},
  \bibinfo{year}{2017}.
\newblock \bibinfo{title}{\emph{Dermatology 2016 Factsheet}}.
\newblock \bibinfo{note}{\url{https://tinyurl.com/y3z39b9r} [Accessed: 15 Jun
  2020]}.
\bibitem[{Dolz et~al.(2020)Dolz, Desrosiers, Wang, Yuan, Shen and
  Ayed}]{dolz2020deep}
\bibinfo{author}{Dolz, J.}, \bibinfo{author}{Desrosiers, C.},
  \bibinfo{author}{Wang, L.}, \bibinfo{author}{Yuan, J.},
  \bibinfo{author}{Shen, D.}, \bibinfo{author}{Ayed, I.B.},
  \bibinfo{year}{2020}.
\newblock \bibinfo{title}{Deep cnn ensembles and suggestive annotations for
  infant brain mri segmentation}.
\newblock \bibinfo{journal}{Computerized Medical Imaging and Graphics}
  \bibinfo{volume}{79}, \bibinfo{pages}{101660}.
\bibitem[{Estava et~al.(2017)Estava, Kuprel, Novoa
  et~al.}]{estava2017dermatologist}
\bibinfo{author}{Estava, A.}, \bibinfo{author}{Kuprel, B.},
  \bibinfo{author}{Novoa, R.}, et~al., \bibinfo{year}{2017}.
\newblock \bibinfo{title}{Dermatologist level classification of skin cancer
  with deep neural networks [j]}.
\newblock \bibinfo{journal}{Nature} \bibinfo{volume}{542},
  \bibinfo{pages}{115}.
\bibitem[{Everingham et~al.(2010)Everingham, Van~Gool, Williams, Winn and
  Zisserman}]{everingham2010pascal}
\bibinfo{author}{Everingham, M.}, \bibinfo{author}{Van~Gool, L.},
  \bibinfo{author}{Williams, C.K.}, \bibinfo{author}{Winn, J.},
  \bibinfo{author}{Zisserman, A.}, \bibinfo{year}{2010}.
\newblock \bibinfo{title}{The pascal visual object classes (voc) challenge}.
\newblock \bibinfo{journal}{International journal of computer vision}
  \bibinfo{volume}{88}, \bibinfo{pages}{303--338}.
\bibitem[{Fujisawa et~al.(2019)Fujisawa, Inoue and
  Nakamura}]{fujisawa2019possibility}
\bibinfo{author}{Fujisawa, Y.}, \bibinfo{author}{Inoue, S.},
  \bibinfo{author}{Nakamura, Y.}, \bibinfo{year}{2019}.
\newblock \bibinfo{title}{The possibility of deep learning-based,
  computer-aided skin tumor classifiers}.
\newblock \bibinfo{journal}{Frontiers in Medicine} \bibinfo{volume}{6},
  \bibinfo{pages}{191}.
\bibitem[{Furey et~al.(2000)Furey, Cristianini, Duffy, Bednarski, Schummer and
  Haussler}]{furey2000support}
\bibinfo{author}{Furey, T.S.}, \bibinfo{author}{Cristianini, N.},
  \bibinfo{author}{Duffy, N.}, \bibinfo{author}{Bednarski, D.W.},
  \bibinfo{author}{Schummer, M.}, \bibinfo{author}{Haussler, D.},
  \bibinfo{year}{2000}.
\newblock \bibinfo{title}{Support vector machine classification and validation
  of cancer tissue samples using microarray expression data}.
\newblock \bibinfo{journal}{Bioinformatics} \bibinfo{volume}{16},
  \bibinfo{pages}{906--914}.
\bibitem[{Garcia-Garcia et~al.(2018)Garcia-Garcia, Orts-Escolano, Oprea,
  Villena-Martinez, Martinez-Gonzalez and Garcia-Rodriguez}]{garcia2018survey}
\bibinfo{author}{Garcia-Garcia, A.}, \bibinfo{author}{Orts-Escolano, S.},
  \bibinfo{author}{Oprea, S.}, \bibinfo{author}{Villena-Martinez, V.},
  \bibinfo{author}{Martinez-Gonzalez, P.}, \bibinfo{author}{Garcia-Rodriguez,
  J.}, \bibinfo{year}{2018}.
\newblock \bibinfo{title}{A survey on deep learning techniques for image and
  video semantic segmentation}.
\newblock \bibinfo{journal}{Applied Soft Computing} \bibinfo{volume}{70},
  \bibinfo{pages}{41--65}.
\bibitem[{Ge et~al.(2017)Ge, Demyanov, Chakravorty, Bowling and
  Garnavi}]{ge2017skin}
\bibinfo{author}{Ge, Z.}, \bibinfo{author}{Demyanov, S.},
  \bibinfo{author}{Chakravorty, R.}, \bibinfo{author}{Bowling, A.},
  \bibinfo{author}{Garnavi, R.}, \bibinfo{year}{2017}.
\newblock \bibinfo{title}{Skin disease recognition using deep saliency features
  and multimodal learning of dermoscopy and clinical images}, in:
  \bibinfo{booktitle}{International Conference on Medical Image Computing and
  Computer-Assisted Intervention}, \bibinfo{organization}{Springer}. pp.
  \bibinfo{pages}{250--258}.
\bibitem[{Gessert et~al.(2020)Gessert, Nielsen, Shaikh, Werner and
  Schlaefer}]{gessert2020skin}
\bibinfo{author}{Gessert, N.}, \bibinfo{author}{Nielsen, M.},
  \bibinfo{author}{Shaikh, M.}, \bibinfo{author}{Werner, R.},
  \bibinfo{author}{Schlaefer, A.}, \bibinfo{year}{2020}.
\newblock \bibinfo{title}{Skin lesion classification using ensembles of
  multi-resolution efficientnets with meta data}.
\newblock \bibinfo{journal}{MethodsX} , \bibinfo{pages}{100864}.
\bibitem[{Ghadiyaram and Bovik(2017)}]{ghadiyaram2017perceptual}
\bibinfo{author}{Ghadiyaram, D.}, \bibinfo{author}{Bovik, A.C.},
  \bibinfo{year}{2017}.
\newblock \bibinfo{title}{Perceptual quality prediction on authentically
  distorted images using a bag of features approach}.
\newblock \bibinfo{journal}{Journal of vision} \bibinfo{volume}{17},
  \bibinfo{pages}{32--32}.
\bibitem[{Glazer and Rigel(2017)}]{glazer2017analysis}
\bibinfo{author}{Glazer, A.M.}, \bibinfo{author}{Rigel, D.S.},
  \bibinfo{year}{2017}.
\newblock \bibinfo{title}{Analysis of trends in geographic distribution of us
  dermatology workforce density}.
\newblock \bibinfo{journal}{JAMA dermatology} \bibinfo{volume}{153},
  \bibinfo{pages}{472--473}.
\bibitem[{Goyal et~al.(2019)Goyal, Oakley, Bansal, Dancey and
  Yap}]{goyal2019skin}
\bibinfo{author}{Goyal, M.}, \bibinfo{author}{Oakley, A.},
  \bibinfo{author}{Bansal, P.}, \bibinfo{author}{Dancey, D.},
  \bibinfo{author}{Yap, M.H.}, \bibinfo{year}{2019}.
\newblock \bibinfo{title}{Skin lesion segmentation in dermoscopic images with
  ensemble deep learning methods}.
\newblock \bibinfo{journal}{IEEE Access} .
\bibitem[{Guo et~al.(2016)Guo, Liu, Oerlemans, Lao, Wu and Lew}]{guo2016deep}
\bibinfo{author}{Guo, Y.}, \bibinfo{author}{Liu, Y.},
  \bibinfo{author}{Oerlemans, A.}, \bibinfo{author}{Lao, S.},
  \bibinfo{author}{Wu, S.}, \bibinfo{author}{Lew, M.S.}, \bibinfo{year}{2016}.
\newblock \bibinfo{title}{Deep learning for visual understanding: A review}.
\newblock \bibinfo{journal}{Neurocomputing} \bibinfo{volume}{187},
  \bibinfo{pages}{27--48}.
\bibitem[{Gutman et~al.(2016)Gutman, Codella, Celebi, Helba, Marchetti, Mishra
  and Halpern}]{gutman2016skin}
\bibinfo{author}{Gutman, D.}, \bibinfo{author}{Codella, N.C.},
  \bibinfo{author}{Celebi, E.}, \bibinfo{author}{Helba, B.},
  \bibinfo{author}{Marchetti, M.}, \bibinfo{author}{Mishra, N.},
  \bibinfo{author}{Halpern, A.}, \bibinfo{year}{2016}.
\newblock \bibinfo{title}{Skin lesion analysis toward melanoma detection: A
  challenge at the international symposium on biomedical imaging (isbi) 2016,
  hosted by the international skin imaging collaboration (isic)}.
\newblock \bibinfo{journal}{arXiv:1605.01397} .
\bibitem[{Ha et~al.(2020)Ha, Liu and Liu}]{ha2020identifying}
\bibinfo{author}{Ha, Q.}, \bibinfo{author}{Liu, B.}, \bibinfo{author}{Liu, F.},
  \bibinfo{year}{2020}.
\newblock \bibinfo{title}{Identifying melanoma images using efficientnet
  ensemble: Winning solution to the siim-isic melanoma classification
  challenge}.
\newblock \bibinfo{journal}{arXiv preprint arXiv:2010.05351} .
\bibitem[{Hameed et~al.(2020)Hameed, Shabut, Ghosh and
  Hossain}]{hameed2020multi}
\bibinfo{author}{Hameed, N.}, \bibinfo{author}{Shabut, A.M.},
  \bibinfo{author}{Ghosh, M.K.}, \bibinfo{author}{Hossain, M.},
  \bibinfo{year}{2020}.
\newblock \bibinfo{title}{Multi-class multi-level classification algorithm for
  skin lesions classification using machine learning techniques}.
\newblock \bibinfo{journal}{Expert Systems with Applications}
  \bibinfo{volume}{141}, \bibinfo{pages}{112961}.
\bibitem[{Harangi(2018)}]{harangi2018skin}
\bibinfo{author}{Harangi, B.}, \bibinfo{year}{2018}.
\newblock \bibinfo{title}{Skin lesion classification with ensembles of deep
  convolutional neural networks}.
\newblock \bibinfo{journal}{Journal of biomedical informatics}
  \bibinfo{volume}{86}, \bibinfo{pages}{25--32}.
\bibitem[{Hasan et~al.(2021a)Hasan, Alam, Elahi, Roy and
  Mart{\'\i}}]{hasan2021drnet}
\bibinfo{author}{Hasan, M.K.}, \bibinfo{author}{Alam, M.A.},
  \bibinfo{author}{Elahi, M.T.E.}, \bibinfo{author}{Roy, S.},
  \bibinfo{author}{Mart{\'\i}, R.}, \bibinfo{year}{2021}a.
\newblock \bibinfo{title}{Drnet: Segmentation and localization of optic disc
  and fovea from diabetic retinopathy image}.
\newblock \bibinfo{journal}{Artificial Intelligence in Medicine}
  \bibinfo{volume}{111}, \bibinfo{pages}{102001}.
\bibitem[{Hasan et~al.(2021b)Hasan, Calvet, Rabbani and
  Bartoli}]{hasan2021detection}
\bibinfo{author}{Hasan, M.K.}, \bibinfo{author}{Calvet, L.},
  \bibinfo{author}{Rabbani, N.}, \bibinfo{author}{Bartoli, A.},
  \bibinfo{year}{2021}b.
\newblock \bibinfo{title}{Detection, segmentation, and 3d pose estimation of
  surgical tools using convolutional neural networks and algebraic geometry}.
\newblock \bibinfo{journal}{Medical Image Analysis} , \bibinfo{pages}{101994}.
\bibitem[{Hasan et~al.(2020)Hasan, Dahal, Samarakoon, Tushar and
  Mart{\'\i}}]{hasan2020dsnet}
\bibinfo{author}{Hasan, M.K.}, \bibinfo{author}{Dahal, L.},
  \bibinfo{author}{Samarakoon, P.N.}, \bibinfo{author}{Tushar, F.I.},
  \bibinfo{author}{Mart{\'\i}, R.}, \bibinfo{year}{2020}.
\newblock \bibinfo{title}{Dsnet: Automatic dermoscopic skin lesion
  segmentation}.
\newblock \bibinfo{journal}{Computers in Biology and Medicine} ,
  \bibinfo{pages}{103738}.
\bibitem[{Hawas et~al.(2020)Hawas, Guo, Du, Polat and Ashour}]{hawas2020oce}
\bibinfo{author}{Hawas, A.R.}, \bibinfo{author}{Guo, Y.}, \bibinfo{author}{Du,
  C.}, \bibinfo{author}{Polat, K.}, \bibinfo{author}{Ashour, A.S.},
  \bibinfo{year}{2020}.
\newblock \bibinfo{title}{Oce-ngc: A neutrosophic graph cut algorithm using
  optimized clustering estimation algorithm for dermoscopic skin lesion
  segmentation}.
\newblock \bibinfo{journal}{Applied Soft Computing} \bibinfo{volume}{86},
  \bibinfo{pages}{105931}.
\bibitem[{He and Garcia(2009)}]{he2009learning}
\bibinfo{author}{He, H.}, \bibinfo{author}{Garcia, E.A.}, \bibinfo{year}{2009}.
\newblock \bibinfo{title}{Learning from imbalanced data}.
\newblock \bibinfo{journal}{IEEE Transactions on knowledge and data
  engineering} \bibinfo{volume}{21}, \bibinfo{pages}{1263--1284}.
\bibitem[{He et~al.(2017)He, Gkioxari, Doll{\'a}r and Girshick}]{he2017mask}
\bibinfo{author}{He, K.}, \bibinfo{author}{Gkioxari, G.},
  \bibinfo{author}{Doll{\'a}r, P.}, \bibinfo{author}{Girshick, R.},
  \bibinfo{year}{2017}.
\newblock \bibinfo{title}{Mask r-cnn}, in: \bibinfo{booktitle}{Proceedings of
  the IEEE international conference on computer vision}, pp.
  \bibinfo{pages}{2961--2969}.
\bibitem[{He et~al.(2015)He, Zhang, Ren and Sun}]{henormal}
\bibinfo{author}{He, K.}, \bibinfo{author}{Zhang, X.}, \bibinfo{author}{Ren,
  S.}, \bibinfo{author}{Sun, J.}, \bibinfo{year}{2015}.
\newblock \bibinfo{title}{Delving deep into rectifiers: Surpassing human-level
  performance on imagenet classification}, in: \bibinfo{booktitle}{Proceedings
  of the IEEE international conference on computer vision}, pp.
  \bibinfo{pages}{1026--1034}.
\bibitem[{He et~al.(2016)He, Zhang, Ren and Sun}]{he2016deep}
\bibinfo{author}{He, K.}, \bibinfo{author}{Zhang, X.}, \bibinfo{author}{Ren,
  S.}, \bibinfo{author}{Sun, J.}, \bibinfo{year}{2016}.
\newblock \bibinfo{title}{Deep residual learning for image recognition}, in:
  \bibinfo{booktitle}{Proceedings of the IEEE conference on computer vision and
  pattern recognition}, pp. \bibinfo{pages}{770--778}.
\bibitem[{Hu et~al.(2018)Hu, Shen and Sun}]{hu2018squeeze}
\bibinfo{author}{Hu, J.}, \bibinfo{author}{Shen, L.}, \bibinfo{author}{Sun,
  G.}, \bibinfo{year}{2018}.
\newblock \bibinfo{title}{Squeeze-and-excitation networks}, in:
  \bibinfo{booktitle}{Proceedings of the IEEE conference on computer vision and
  pattern recognition}, pp. \bibinfo{pages}{7132--7141}.
\bibitem[{Huang et~al.(2017)Huang, Liu, Van Der~Maaten and
  Weinberger}]{huang2017densely}
\bibinfo{author}{Huang, G.}, \bibinfo{author}{Liu, Z.}, \bibinfo{author}{Van
  Der~Maaten, L.}, \bibinfo{author}{Weinberger, K.Q.}, \bibinfo{year}{2017}.
\newblock \bibinfo{title}{Densely connected convolutional networks}, in:
  \bibinfo{booktitle}{Proceedings of the IEEE conference on computer vision and
  pattern recognition}, pp. \bibinfo{pages}{4700--4708}.
\bibitem[{Ioffe and Szegedyet(2015)}]{Sergey}
\bibinfo{author}{Ioffe, S.}, \bibinfo{author}{Szegedyet, C.},
  \bibinfo{year}{2015}.
\newblock \bibinfo{title}{{Batch Normalization:} accelerating deep network
  training by reducing internal covariate shift}.
\newblock \bibinfo{journal}{arXiv:1502.03167} .
\bibitem[{{ISIC}(2018)}]{ISICArch3}
\bibinfo{author}{{ISIC}}, \bibinfo{year}{2018}.
\newblock \bibinfo{title}{\emph{ISIC Archive}}.
\newblock \bibinfo{note}{\url{https://tinyurl.com/11hgg83u} [Accessed: 09 May
  2020]}.
\bibitem[{Jahanifar et~al.(2018)Jahanifar, Tajeddin, Asl and
  Gooya}]{jahanifar2018supervised}
\bibinfo{author}{Jahanifar, M.}, \bibinfo{author}{Tajeddin, N.Z.},
  \bibinfo{author}{Asl, B.M.}, \bibinfo{author}{Gooya, A.},
  \bibinfo{year}{2018}.
\newblock \bibinfo{title}{Supervised saliency map driven segmentation of
  lesions in dermoscopic images}.
\newblock \bibinfo{journal}{IEEE journal of biomedical and health informatics}
  \bibinfo{volume}{23}, \bibinfo{pages}{509--518}.
\bibitem[{Kermany et~al.(2018)Kermany, Goldbaum, Cai, Valentim, Liang, Baxter,
  McKeown, Yang, Wu, Yan et~al.}]{kermany2018identifying}
\bibinfo{author}{Kermany, D.S.}, \bibinfo{author}{Goldbaum, M.},
  \bibinfo{author}{Cai, W.}, \bibinfo{author}{Valentim, C.C.},
  \bibinfo{author}{Liang, H.}, \bibinfo{author}{Baxter, S.L.},
  \bibinfo{author}{McKeown, A.}, \bibinfo{author}{Yang, G.},
  \bibinfo{author}{Wu, X.}, \bibinfo{author}{Yan, F.}, et~al.,
  \bibinfo{year}{2018}.
\newblock \bibinfo{title}{Identifying medical diagnoses and treatable diseases
  by image-based deep learning}.
\newblock \bibinfo{journal}{Cell} \bibinfo{volume}{172},
  \bibinfo{pages}{1122--1131}.
\bibitem[{Khan et~al.(2020)Khan, Sharif, Akram, Bukhari and
  Nayak}]{khan2020developed}
\bibinfo{author}{Khan, M.A.}, \bibinfo{author}{Sharif, M.},
  \bibinfo{author}{Akram, T.}, \bibinfo{author}{Bukhari, S.A.C.},
  \bibinfo{author}{Nayak, R.S.}, \bibinfo{year}{2020}.
\newblock \bibinfo{title}{Developed newton-raphson based deep features
  selection framework for skin lesion recognition}.
\newblock \bibinfo{journal}{Pattern Recognition Letters} \bibinfo{volume}{129},
  \bibinfo{pages}{293--303}.
\bibitem[{Krizhevsky et~al.(2012)Krizhevsky, Sutskever and
  Hinton}]{krizhevsky2012imagenet}
\bibinfo{author}{Krizhevsky, A.}, \bibinfo{author}{Sutskever, I.},
  \bibinfo{author}{Hinton, G.E.}, \bibinfo{year}{2012}.
\newblock \bibinfo{title}{Imagenet classification with deep convolutional
  neural networks}, in: \bibinfo{booktitle}{Advances in neural information
  processing systems}, pp. \bibinfo{pages}{1097--1105}.
\bibitem[{Kumar et~al.(2016)Kumar, Kim, Lyndon, Fulham and
  Feng}]{kumar2016ensemble}
\bibinfo{author}{Kumar, A.}, \bibinfo{author}{Kim, J.},
  \bibinfo{author}{Lyndon, D.}, \bibinfo{author}{Fulham, M.},
  \bibinfo{author}{Feng, D.}, \bibinfo{year}{2016}.
\newblock \bibinfo{title}{An ensemble of fine-tuned convolutional neural
  networks for medical image classification}.
\newblock \bibinfo{journal}{IEEE journal of biomedical and health informatics}
  \bibinfo{volume}{21}, \bibinfo{pages}{31--40}.
\bibitem[{Lattoofi et~al.(2019)Lattoofi, Al-sharuee, Kamil, Obaid, Mahidi, Omar
  et~al.}]{lattoofi2019melanoma}
\bibinfo{author}{Lattoofi, N.F.}, \bibinfo{author}{Al-sharuee, I.F.},
  \bibinfo{author}{Kamil, M.Y.}, \bibinfo{author}{Obaid, A.H.},
  \bibinfo{author}{Mahidi, A.A.}, \bibinfo{author}{Omar, A.A.}, et~al.,
  \bibinfo{year}{2019}.
\newblock \bibinfo{title}{Melanoma skin cancer detection based on abcd rule},
  in: \bibinfo{booktitle}{2019 First International Conference of Computer and
  Applied Sciences (CAS)}, \bibinfo{organization}{IEEE}. pp.
  \bibinfo{pages}{154--157}.
\bibitem[{Lee et~al.(2018)Lee, Jung and Won}]{lee2018wonderm}
\bibinfo{author}{Lee, Y.C.}, \bibinfo{author}{Jung, S.H.},
  \bibinfo{author}{Won, H.H.}, \bibinfo{year}{2018}.
\newblock \bibinfo{title}{Wonderm: Skin lesion classification with fine-tuned
  neural networks}.
\newblock \bibinfo{journal}{arXiv preprint arXiv:1808.03426} .
\bibitem[{Lin et~al.(2013)Lin, Chen and Yan}]{lin2013network}
\bibinfo{author}{Lin, M.}, \bibinfo{author}{Chen, Q.}, \bibinfo{author}{Yan,
  S.}, \bibinfo{year}{2013}.
\newblock \bibinfo{title}{Network in network}.
\newblock \bibinfo{journal}{arXiv:1312.4400} .
\bibitem[{Long et~al.(2015)Long, Shelhamer and Darrell}]{long2015fully}
\bibinfo{author}{Long, J.}, \bibinfo{author}{Shelhamer, E.},
  \bibinfo{author}{Darrell, T.}, \bibinfo{year}{2015}.
\newblock \bibinfo{title}{Fully convolutional networks for semantic
  segmentation}, in: \bibinfo{booktitle}{Proceedings of the IEEE conference on
  computer vision and pattern recognition}, pp. \bibinfo{pages}{3431--3440}.
\bibitem[{Lowe(2004)}]{lowe2004distinctive}
\bibinfo{author}{Lowe, D.G.}, \bibinfo{year}{2004}.
\newblock \bibinfo{title}{Distinctive image features from scale-invariant
  keypoints}.
\newblock \bibinfo{journal}{International journal of computer vision}
  \bibinfo{volume}{60}, \bibinfo{pages}{91--110}.
\bibitem[{Ma{\'c}kiewicz and Ratajczak(1993)}]{mackiewicz1993principal}
\bibinfo{author}{Ma{\'c}kiewicz, A.}, \bibinfo{author}{Ratajczak, W.},
  \bibinfo{year}{1993}.
\newblock \bibinfo{title}{Principal components analysis (pca)}.
\newblock \bibinfo{journal}{Computers \& Geosciences} \bibinfo{volume}{19},
  \bibinfo{pages}{303--342}.
\bibitem[{Maglogiannis and Doukas(2009)}]{maglogiannis2009overview}
\bibinfo{author}{Maglogiannis, I.}, \bibinfo{author}{Doukas, C.N.},
  \bibinfo{year}{2009}.
\newblock \bibinfo{title}{Overview of advanced computer vision systems for skin
  lesions characterization}.
\newblock \bibinfo{journal}{IEEE transactions on information technology in
  biomedicine} \bibinfo{volume}{13}, \bibinfo{pages}{721--733}.
\bibitem[{Mahbod et~al.(2019)Mahbod, Schaefer, Ellinger, Ecker, Pitiot and
  Wang}]{mahbod2019fusing}
\bibinfo{author}{Mahbod, A.}, \bibinfo{author}{Schaefer, G.},
  \bibinfo{author}{Ellinger, I.}, \bibinfo{author}{Ecker, R.},
  \bibinfo{author}{Pitiot, A.}, \bibinfo{author}{Wang, C.},
  \bibinfo{year}{2019}.
\newblock \bibinfo{title}{Fusing fine-tuned deep features for skin lesion
  classification}.
\newblock \bibinfo{journal}{Computerized Medical Imaging and Graphics}
  \bibinfo{volume}{71}, \bibinfo{pages}{19--29}.
\bibitem[{Mahbod et~al.(2020)Mahbod, Schaefer, Wang, Dorffner, Ecker and
  Ellinger}]{mahbod2020transfer}
\bibinfo{author}{Mahbod, A.}, \bibinfo{author}{Schaefer, G.},
  \bibinfo{author}{Wang, C.}, \bibinfo{author}{Dorffner, G.},
  \bibinfo{author}{Ecker, R.}, \bibinfo{author}{Ellinger, I.},
  \bibinfo{year}{2020}.
\newblock \bibinfo{title}{Transfer learning using a multi-scale and
  multi-network ensemble for skin lesion classification}.
\newblock \bibinfo{journal}{Computer methods and programs in biomedicine}
  \bibinfo{volume}{193}, \bibinfo{pages}{105475}.
\bibitem[{Moitra and Mandal(2020)}]{moitra2020prediction}
\bibinfo{author}{Moitra, D.}, \bibinfo{author}{Mandal, R.K.},
  \bibinfo{year}{2020}.
\newblock \bibinfo{title}{Prediction of non-small cell lung cancer histology by
  a deep ensemble of convolutional and bidirectional recurrent neural network}.
\newblock \bibinfo{journal}{Journal of Digital Imaging} ,
  \bibinfo{pages}{1--8}.
\bibitem[{Mporas et~al.(2020)Mporas, Perikos and Paraskevas}]{mporas2020color}
\bibinfo{author}{Mporas, I.}, \bibinfo{author}{Perikos, I.},
  \bibinfo{author}{Paraskevas, M.}, \bibinfo{year}{2020}.
\newblock \bibinfo{title}{Color models for skin lesion classification from
  dermatoscopic images}, in: \bibinfo{booktitle}{Advances in Integrations of
  Intelligent Methods}. \bibinfo{publisher}{Springer}, pp.
  \bibinfo{pages}{85--98}.
\bibitem[{Nachbar et~al.(1994)Nachbar, Stolz, Merkle, Cognetta, Vogt,
  Landthaler, Bilek, Braun-Falco and Plewig}]{nachbar1994abcd}
\bibinfo{author}{Nachbar, F.}, \bibinfo{author}{Stolz, W.},
  \bibinfo{author}{Merkle, T.}, \bibinfo{author}{Cognetta, A.B.},
  \bibinfo{author}{Vogt, T.}, \bibinfo{author}{Landthaler, M.},
  \bibinfo{author}{Bilek, P.}, \bibinfo{author}{Braun-Falco, O.},
  \bibinfo{author}{Plewig, G.}, \bibinfo{year}{1994}.
\newblock \bibinfo{title}{The abcd rule of dermatoscopy: high prospective value
  in the diagnosis of doubtful melanocytic skin lesions}.
\newblock \bibinfo{journal}{Journal of the American Academy of Dermatology}
  \bibinfo{volume}{30}, \bibinfo{pages}{551--559}.
\bibitem[{Navarro et~al.(2018)Navarro, Escudero-Vi{\~n}olo and
  Besc{\'o}s}]{navarro2018accurate}
\bibinfo{author}{Navarro, F.}, \bibinfo{author}{Escudero-Vi{\~n}olo, M.},
  \bibinfo{author}{Besc{\'o}s, J.}, \bibinfo{year}{2018}.
\newblock \bibinfo{title}{Accurate segmentation and registration of skin lesion
  images to evaluate lesion change}.
\newblock \bibinfo{journal}{IEEE journal of biomedical and health informatics}
  \bibinfo{volume}{23}, \bibinfo{pages}{501--508}.
\bibitem[{Odena et~al.(2016)Odena, Dumoulin and Olah}]{odena2016deconvolution}
\bibinfo{author}{Odena, A.}, \bibinfo{author}{Dumoulin, V.},
  \bibinfo{author}{Olah, C.}, \bibinfo{year}{2016}.
\newblock \bibinfo{title}{Deconvolution and checkerboard artifacts}.
\newblock \bibinfo{journal}{Distill} \bibinfo{volume}{1}, \bibinfo{pages}{e3}.
\bibitem[{Oliveira et~al.(2018)Oliveira, Papa, Pereira and
  Tavares}]{oliveira2018computational}
\bibinfo{author}{Oliveira, R.B.}, \bibinfo{author}{Papa, J.P.},
  \bibinfo{author}{Pereira, A.S.}, \bibinfo{author}{Tavares, J.M.R.},
  \bibinfo{year}{2018}.
\newblock \bibinfo{title}{Computational methods for pigmented skin lesion
  classification in images: review and future trends}.
\newblock \bibinfo{journal}{Neural Computing and Applications}
  \bibinfo{volume}{29}, \bibinfo{pages}{613--636}.
\bibitem[{{\"O}zt{\"u}rk and {\"O}zkaya(2020)}]{ozturk2020skin}
\bibinfo{author}{{\"O}zt{\"u}rk, {\c{S}}.}, \bibinfo{author}{{\"O}zkaya, U.},
  \bibinfo{year}{2020}.
\newblock \bibinfo{title}{Skin lesion segmentation with improved convolutional
  neural network}.
\newblock \bibinfo{journal}{Journal of digital imaging} .
\bibitem[{Pacheco et~al.(2019)Pacheco, Ali and Trappenberg}]{pacheco2019skin}
\bibinfo{author}{Pacheco, A.G.}, \bibinfo{author}{Ali, A.R.},
  \bibinfo{author}{Trappenberg, T.}, \bibinfo{year}{2019}.
\newblock \bibinfo{title}{Skin cancer detection based on deep learning and
  entropy to detect outlier samples}.
\newblock \bibinfo{journal}{arXiv preprint arXiv:1909.04525} .
\bibitem[{Pour and Seker(2020)}]{pour2020transform}
\bibinfo{author}{Pour, M.P.}, \bibinfo{author}{Seker, H.},
  \bibinfo{year}{2020}.
\newblock \bibinfo{title}{Transform domain representation-driven convolutional
  neural networks for skin lesion segmentation}.
\newblock \bibinfo{journal}{Expert Systems with Applications}
  \bibinfo{volume}{144}, \bibinfo{pages}{113129}.
\bibitem[{Qin et~al.(2020)Qin, Liu, Zhu and Xue}]{qin2020gan}
\bibinfo{author}{Qin, Z.}, \bibinfo{author}{Liu, Z.}, \bibinfo{author}{Zhu,
  P.}, \bibinfo{author}{Xue, Y.}, \bibinfo{year}{2020}.
\newblock \bibinfo{title}{A gan-based image synthesis method for skin lesion
  classification}.
\newblock \bibinfo{journal}{Computer Methods and Programs in Biomedicine} ,
  \bibinfo{pages}{105568}.
\bibitem[{Rajpurkar et~al.(2017)Rajpurkar, Irvin, Zhu, Yang, Mehta, Duan, Ding,
  Bagul, Langlotz, Shpanskaya et~al.}]{rajpurkar2017chexnet}
\bibinfo{author}{Rajpurkar, P.}, \bibinfo{author}{Irvin, J.},
  \bibinfo{author}{Zhu, K.}, \bibinfo{author}{Yang, B.},
  \bibinfo{author}{Mehta, H.}, \bibinfo{author}{Duan, T.},
  \bibinfo{author}{Ding, D.}, \bibinfo{author}{Bagul, A.},
  \bibinfo{author}{Langlotz, C.}, \bibinfo{author}{Shpanskaya, K.}, et~al.,
  \bibinfo{year}{2017}.
\newblock \bibinfo{title}{Chexnet: Radiologist-level pneumonia detection on
  chest x-rays with deep learning}.
\newblock \bibinfo{journal}{arXiv:1711.05225} .
\bibitem[{Ronneberger et~al.(2015)Ronneberger, Fischer and
  Brox}]{ronneberger2015u}
\bibinfo{author}{Ronneberger, O.}, \bibinfo{author}{Fischer, P.},
  \bibinfo{author}{Brox, T.}, \bibinfo{year}{2015}.
\newblock \bibinfo{title}{U-net: Convolutional networks for biomedical image
  segmentation}, in: \bibinfo{booktitle}{International Conference on Medical
  image computing and computer-assisted intervention},
  \bibinfo{organization}{Springer}. pp. \bibinfo{pages}{234--241}.
\bibitem[{Sarker et~al.(2018)Sarker, Rashwan, Akram, Banu, Saleh, Singh,
  Chowdhury, Abdulwahab, Romani, Radeva et~al.}]{sarker2018slsdeep}
\bibinfo{author}{Sarker, M.M.K.}, \bibinfo{author}{Rashwan, H.A.},
  \bibinfo{author}{Akram, F.}, \bibinfo{author}{Banu, S.F.},
  \bibinfo{author}{Saleh, A.}, \bibinfo{author}{Singh, V.K.},
  \bibinfo{author}{Chowdhury, F.U.}, \bibinfo{author}{Abdulwahab, S.},
  \bibinfo{author}{Romani, S.}, \bibinfo{author}{Radeva, P.}, et~al.,
  \bibinfo{year}{2018}.
\newblock \bibinfo{title}{Slsdeep: Skin lesion segmentation based on dilated
  residual and pyramid pooling networks}, in: \bibinfo{booktitle}{International
  Conference on Medical Image Computing and Computer-Assisted Intervention},
  \bibinfo{organization}{Springer}. pp. \bibinfo{pages}{21--29}.
\bibitem[{Savelli et~al.(2020)Savelli, Bria, Molinara, Marrocco and
  Tortorella}]{savelli2020multi}
\bibinfo{author}{Savelli, B.}, \bibinfo{author}{Bria, A.},
  \bibinfo{author}{Molinara, M.}, \bibinfo{author}{Marrocco, C.},
  \bibinfo{author}{Tortorella, F.}, \bibinfo{year}{2020}.
\newblock \bibinfo{title}{A multi-context cnn ensemble for small lesion
  detection}.
\newblock \bibinfo{journal}{Artificial Intelligence in Medicine}
  \bibinfo{volume}{103}, \bibinfo{pages}{101749}.
\bibitem[{Scheffe(1999)}]{scheffe1999analysis}
\bibinfo{author}{Scheffe, H.}, \bibinfo{year}{1999}.
\newblock \bibinfo{title}{The analysis of variance}.
  volume~\bibinfo{volume}{72}.
\newblock \bibinfo{publisher}{John Wiley \& Sons}.
\bibitem[{Serte and Demirel(2019)}]{serte2019gabor}
\bibinfo{author}{Serte, S.}, \bibinfo{author}{Demirel, H.},
  \bibinfo{year}{2019}.
\newblock \bibinfo{title}{Gabor wavelet-based deep learning for skin lesion
  classification}.
\newblock \bibinfo{journal}{Computers in biology and medicine}
  \bibinfo{volume}{113}, \bibinfo{pages}{103423}.
\bibitem[{Shahin et~al.(2018)Shahin, Kamal and Elattar}]{shahin2018deep}
\bibinfo{author}{Shahin, A.H.}, \bibinfo{author}{Kamal, A.},
  \bibinfo{author}{Elattar, M.A.}, \bibinfo{year}{2018}.
\newblock \bibinfo{title}{Deep ensemble learning for skin lesion classification
  from dermoscopic images}, in: \bibinfo{booktitle}{2018 9th Cairo
  International Biomedical Engineering Conference (CIBEC)},
  \bibinfo{organization}{IEEE}. pp. \bibinfo{pages}{150--153}.
\bibitem[{Siegel et~al.(2020)Siegel, Miller and Jemal}]{siegel2020cancer}
\bibinfo{author}{Siegel, R.L.}, \bibinfo{author}{Miller, K.D.},
  \bibinfo{author}{Jemal, A.}, \bibinfo{year}{2020}.
\newblock \bibinfo{title}{Cancer statistics, 2020}.
\newblock \bibinfo{journal}{CA: A Cancer Journal for Clinicians}
  \bibinfo{volume}{70}, \bibinfo{pages}{7--30}.
\bibitem[{Song et~al.(2020)Song, Lin, Wang and Wang}]{song2020end}
\bibinfo{author}{Song, L.}, \bibinfo{author}{Lin, J.P.}, \bibinfo{author}{Wang,
  Z.J.}, \bibinfo{author}{Wang, H.}, \bibinfo{year}{2020}.
\newblock \bibinfo{title}{An end-to-end multi-task deep learning framework for
  skin lesion analysis}.
\newblock \bibinfo{journal}{IEEE Journal of Biomedical and Health Informatics}
  .
\bibitem[{Srivastava et~al.(2014)Srivastava, Hinton, Krizhevsky, Sutskever and
  Salakhutdinov}]{srivastava2014dropout}
\bibinfo{author}{Srivastava, N.}, \bibinfo{author}{Hinton, G.},
  \bibinfo{author}{Krizhevsky, A.}, \bibinfo{author}{Sutskever, I.},
  \bibinfo{author}{Salakhutdinov, R.}, \bibinfo{year}{2014}.
\newblock \bibinfo{title}{Dropout: a simple way to prevent neural networks from
  overfitting}.
\newblock \bibinfo{journal}{The journal of machine learning research}
  \bibinfo{volume}{15}, \bibinfo{pages}{1929--1958}.
\bibitem[{Szegedy et~al.(2016)Szegedy, Vanhoucke, Ioffe, Shlens and
  Wojna}]{szegedy2016rethinking}
\bibinfo{author}{Szegedy, C.}, \bibinfo{author}{Vanhoucke, V.},
  \bibinfo{author}{Ioffe, S.}, \bibinfo{author}{Shlens, J.},
  \bibinfo{author}{Wojna, Z.}, \bibinfo{year}{2016}.
\newblock \bibinfo{title}{Rethinking the inception architecture for computer
  vision}, in: \bibinfo{booktitle}{Proceedings of the IEEE conference on
  computer vision and pattern recognition}, pp. \bibinfo{pages}{2818--2826}.
\bibitem[{Tan and Le(2019)}]{tan2019efficientnet}
\bibinfo{author}{Tan, M.}, \bibinfo{author}{Le, Q.V.}, \bibinfo{year}{2019}.
\newblock \bibinfo{title}{Efficientnet: Rethinking model scaling for
  convolutional neural networks}.
\newblock \bibinfo{journal}{arXiv preprint arXiv:1905.11946} .
\bibitem[{Tang et~al.(2019)Tang, Yang, Yuan et~al.}]{tang2019multi}
\bibinfo{author}{Tang, Y.}, \bibinfo{author}{Yang, F.}, \bibinfo{author}{Yuan,
  S.}, et~al., \bibinfo{year}{2019}.
\newblock \bibinfo{title}{A multi-stage framework with context information
  fusion structure for skin lesion segmentation}, in: \bibinfo{booktitle}{2019
  IEEE 16th International Symposium on Biomedical Imaging (ISBI 2019)},
  \bibinfo{organization}{IEEE}. pp. \bibinfo{pages}{1407--1410}.
\bibitem[{Torrey and Shavlik(2010)}]{torrey2010transfer}
\bibinfo{author}{Torrey, L.}, \bibinfo{author}{Shavlik, J.},
  \bibinfo{year}{2010}.
\newblock \bibinfo{title}{Transfer learning}, in: \bibinfo{booktitle}{Handbook
  of research on machine learning applications and trends: algorithms, methods,
  and techniques}. \bibinfo{publisher}{IGI Global}, pp.
  \bibinfo{pages}{242--264}.
\bibitem[{Valle et~al.(2020)Valle, Fornaciali, Menegola, Tavares, Bittencourt,
  Li and Avila}]{valle2020data}
\bibinfo{author}{Valle, E.}, \bibinfo{author}{Fornaciali, M.},
  \bibinfo{author}{Menegola, A.}, \bibinfo{author}{Tavares, J.},
  \bibinfo{author}{Bittencourt, F.V.}, \bibinfo{author}{Li, L.T.},
  \bibinfo{author}{Avila, S.}, \bibinfo{year}{2020}.
\newblock \bibinfo{title}{Data, depth, and design: Learning reliable models for
  skin lesion analysis}.
\newblock \bibinfo{journal}{Neurocomputing} \bibinfo{volume}{383},
  \bibinfo{pages}{303--313}.
\bibitem[{Venugopal and Stoffel(2019)}]{venugopal2019colorectal}
\bibinfo{author}{Venugopal, A.}, \bibinfo{author}{Stoffel, E.M.},
  \bibinfo{year}{2019}.
\newblock \bibinfo{title}{Colorectal cancer in young adults}.
\newblock \bibinfo{journal}{Current treatment options in gastroenterology}
  \bibinfo{volume}{17}, \bibinfo{pages}{89--98}.
\bibitem[{Xie et~al.(2020a)Xie, Yang, Liu, Jiang, Zheng and Wang}]{xie2020skin}
\bibinfo{author}{Xie, F.}, \bibinfo{author}{Yang, J.}, \bibinfo{author}{Liu,
  J.}, \bibinfo{author}{Jiang, Z.}, \bibinfo{author}{Zheng, Y.},
  \bibinfo{author}{Wang, Y.}, \bibinfo{year}{2020}a.
\newblock \bibinfo{title}{Skin lesion segmentation using high-resolution
  convolutional neural network}.
\newblock \bibinfo{journal}{Computer Methods and Programs in Biomedicine}
  \bibinfo{volume}{186}, \bibinfo{pages}{105241}.
\bibitem[{Xie et~al.(2020b)Xie, Zhang, Xia and Shen}]{xie2020mutual}
\bibinfo{author}{Xie, Y.}, \bibinfo{author}{Zhang, J.}, \bibinfo{author}{Xia,
  Y.}, \bibinfo{author}{Shen, C.}, \bibinfo{year}{2020}b.
\newblock \bibinfo{title}{A mutual bootstrapping model for automated skin
  lesion segmentation and classification}.
\newblock \bibinfo{journal}{IEEE Transactions on Medical Imaging} .
\bibitem[{Yadav and Jadhav(2019)}]{yadav2019deep}
\bibinfo{author}{Yadav, S.S.}, \bibinfo{author}{Jadhav, S.M.},
  \bibinfo{year}{2019}.
\newblock \bibinfo{title}{Deep convolutional neural network based medical image
  classification for disease diagnosis}.
\newblock \bibinfo{journal}{Journal of Big Data} \bibinfo{volume}{6},
  \bibinfo{pages}{113}.
\bibitem[{Yilmaz and Trocan(2020)}]{yilmaz2020benign}
\bibinfo{author}{Yilmaz, E.}, \bibinfo{author}{Trocan, M.},
  \bibinfo{year}{2020}.
\newblock \bibinfo{title}{Benign and malignant skin lesion classification
  comparison for three deep-learning architectures}, in:
  \bibinfo{booktitle}{Asian Conference on Intelligent Information and Database
  Systems}, \bibinfo{organization}{Springer}. pp. \bibinfo{pages}{514--524}.
\bibitem[{Yu et~al.(2020)Yu, Jiang, Zhou, He, Ni, Chen, Wang and
  Lei}]{yu2020convolutional}
\bibinfo{author}{Yu, Z.}, \bibinfo{author}{Jiang, F.}, \bibinfo{author}{Zhou,
  F.}, \bibinfo{author}{He, X.}, \bibinfo{author}{Ni, D.},
  \bibinfo{author}{Chen, S.}, \bibinfo{author}{Wang, T.}, \bibinfo{author}{Lei,
  B.}, \bibinfo{year}{2020}.
\newblock \bibinfo{title}{Convolutional descriptors aggregation via cross-net
  for skin lesion recognition}.
\newblock \bibinfo{journal}{Applied Soft Computing} , \bibinfo{pages}{106281}.
\bibitem[{Yuan(2017)}]{yuan2017automatic}
\bibinfo{author}{Yuan, Y.}, \bibinfo{year}{2017}.
\newblock \bibinfo{title}{Automatic skin lesion segmentation with fully
  convolutional-deconvolutional networks}.
\newblock \bibinfo{journal}{arXiv preprint arXiv:1703.05165} .
\bibitem[{Zhang et~al.(2019)Zhang, Xie, Xia and Shen}]{zhang2019attention}
\bibinfo{author}{Zhang, J.}, \bibinfo{author}{Xie, Y.}, \bibinfo{author}{Xia,
  Y.}, \bibinfo{author}{Shen, C.}, \bibinfo{year}{2019}.
\newblock \bibinfo{title}{Attention residual learning for skin lesion
  classification}.
\newblock \bibinfo{journal}{IEEE transactions on medical imaging}
  \bibinfo{volume}{38}, \bibinfo{pages}{2092--2103}.
\bibitem[{Zhang et~al.(2020)Zhang, Cai, Wang, Tian, Wang and
  Badami}]{zhang2020skin}
\bibinfo{author}{Zhang, N.}, \bibinfo{author}{Cai, Y.X.},
  \bibinfo{author}{Wang, Y.Y.}, \bibinfo{author}{Tian, Y.T.},
  \bibinfo{author}{Wang, X.L.}, \bibinfo{author}{Badami, B.},
  \bibinfo{year}{2020}.
\newblock \bibinfo{title}{Skin cancer diagnosis based on optimized
  convolutional neural network}.
\newblock \bibinfo{journal}{Artificial Intelligence in Medicine}
  \bibinfo{volume}{102}, \bibinfo{pages}{101756}.

\end{thebibliography}

\end{document}